%% file: main.tex
\documentclass[manuscript,screen]{acmart}

\AtBeginDocument{%
  }

\setcopyright{acmlicensed}
\acmJournal{TOIS}
\acmYear{2025} \acmVolume{1} \acmNumber{1} \acmArticle{1} \acmMonth{1}\acmDOI{10.1145/3716320}

\acmConference{ACM Transactions on Information Systems}{July 03--05,
  2024}{Woodstock, NY}
\usepackage{graphicx}
\usepackage{float}
\usepackage{subfigure}
\usepackage{wrapfig}
\usepackage{float}
\usepackage[utf8]{inputenc} % allow utf-8 input
\usepackage[T1]{fontenc}    % use 8-bit T1 fonts
\usepackage{hyperref}       % hyperlinks
\usepackage{url}            % simple URL typesetting
\usepackage{booktabs}       % professional-quality tables
\usepackage{amsfonts}       % blackboard math symbols
\usepackage{nicefrac}       % compact symbols for 1/2, etc.
\usepackage{microtype}      % microtypography
\usepackage{xcolor}         % colors
\usepackage{ragged2e} 
\usepackage{booktabs,makecell, multirow, tabularx}
\usepackage{colortbl}
\usepackage{algorithmic}
\usepackage[ruled,linesnumbered,vlined]{algorithm2e}
\usepackage{enumitem}
\usepackage{verbatim}
\usepackage{CJKutf8}

\newcommand{\Mat}[1]{\mathbf{#1}}

\newcommand{\ie}{\textit{i.e., }}
\newcommand{\eg}{\textit{e.g., }}

\definecolor{mygray}{rgb}{0.9, 0.9, 0.9}
\usepackage{tcolorbox} % for colored boxes
\usepackage{graphicx} % for including images
\tcbuselibrary{skins,breakable} % for tcolorbox skins 

\begin{document}

\title{Reinforced Prompt Personalization for Recommendation with Large Language Models}

\author{Wenyu Mao}
\email{wenyumao2@gmail.com}
\orcid{0009-0003-1348-8412}
\affiliation{%
  \institution{University of Science and Technology of China}
  \city{Heifei}
  \country{China}
}

\author{Jiancan Wu*}\thanks{* corresponding author}
\affiliation{%
  \institution{University of Science and Technology of China}
  \city{Hefei}
  \country{China}}
\email{wujcan@gmail.com}
\orcid{0000-0002-6941-5218}

\author{Weijian Chen}
\affiliation{%
  \institution{Institute of Dataspace, Hefei Comprehensive National Science Center}
  \city{Hefei}
  \country{China}
}\email{naure@mail.ustc.edu.cn}
\orcid{0000-0002-4752-2629}

\author{Chongming Gao}
\affiliation{%
 \institution{University of Science and Technology of China}
 \city{Hefei}
 \country{China}}
\email{	chongming.gao@gmail.com}
\orcid{0000-0002-5187-9196}

\author{Xiang Wang}
\affiliation{%
  \institution{MoE Key Laboratory of Brain-inspired Intelligent Perception and Cognition, University of Science
and Technology of China}
  \city{Hefei}
  \country{China}}
\email{	xiangwang@u.nus.edu}
\orcid{0000-0002-6148-6329}

\author{Xiangnan He}
\affiliation{%
  \institution{MoE Key Laboratory of Brain-inspired Intelligent Perception and Cognition, University of Science
and Technology of China, China}
  \city{Hefei}
  \country{China}}
\email{xiangnanhe@gmail.com}
\orcid{0000-0001-8472-7992}

\renewcommand{\shortauthors}{Wenyu Mao et al.}

\begin{abstract}
% Large Language Models (LLMs) have showcased exceptional capabilities in intent understanding and knowledge utilization. These
% advancements bring exciting opportunities for LLMs to comprehend user preferences and provide personalized recommendations. To unlock the potential of pre-trained LLMs in recommendation, recent studies have drawn increasing attention to the role
% of discrete prompts. 
Designing effective prompts can empower LLMs to understand user preferences and provide recommendations with
intent comprehension and knowledge utilization capabilities. Nevertheless, recent studies predominantly concentrate on task-wise prompting, developing fixed prompt templates shared across all users in a given recommendation task (\eg rating or ranking). Although convenient, task-wise prompting overlooks individual user differences, leading to inaccurate analysis of user interests. In this work, we introduce the concept of instance-wise prompting, aiming at personalizing discrete prompts for individual users. Toward this end, we propose Reinforced Prompt Personalization (RPP) to realize it automatically. To improve efficiency and quality, RPP personalizes prompts at the sentence level rather than searching in the vast vocabulary word-by-word. Specifically, RPP breaks down the prompt into four patterns, tailoring patterns based on multi-agent and combining them. Then the personalized prompts interact with LLMs (environment) iteratively, to boost LLMs' recommending performance (reward). In addition to RPP, to improve the scalability of action space, our proposal of RPP+ dynamically refines the selected actions with LLMs throughout the iterative process. Extensive experiments on various datasets demonstrate the superiority of RPP/RPP+ over traditional recommender models, few-shot methods, and other prompt-based methods, underscoring the significance of instance-wise prompting in LLMs for recommendation. Our code is available at \url{https://github.com/maowenyu-11/RPP}.
\end{abstract}






\maketitle

\input{1_introduction.tex}
\input{3_problem_formulation.tex}

\input{4_methodology.tex}
\input{5_experiments.tex}
\input{2_related_work}

\input{6_conclusion.tex}
\bibliographystyle{ACM-Reference-Format}
\bibliography{main}


\end{document}

%% file: 1_introduction.tex
\section{Introduction}
Recommender systems (RSs) are widely applied in real life to offer personalized item recommendations that align with user preference. Recently, large language models (LLMs), such as ChatGPT, GPT4, PaLM-E, and LLaMA, have demonstrated exceptional capabilities in semantic understanding, intent reasoning, and knowledge utilization~\cite{zhao2023survey}. These advancements empower LLMs to address recommendation tasks by modeling users' behavior into natural language, attracting considerable attention~\cite{li2023preliminary,geng2022recommendation,iclr_ding}.

To unleash the recommendation capabilities of LLMs, 
a promising solution lies in tailoring prompts for the specific recommendation task~\cite{zhang2023recommendation,geng2022recommendation},
beyond fine-tuning the massive parameters~\cite{bao2023tallrec} that can be resource-hungry.
% as fine-tuning by updating the massive parameters can be costly.
Existing methods~\cite{wu2023survey,gao2023chat,hou2023large} have investigated the \textbf{task-wise prompting}, dedicated to establishing fixed prompt templates shared across all users.
To summarize, most of these task-wise prompts for recommendation tasks include the following patterns:
\begin{itemize}[leftmargin=*]
\item \textit{Role-playing} assigns a specific role for LLMs to play~\cite{rolerec,toolrec} in the recommendation tasks, enabling LLMs to respond with domain knowledge aligned to that role.
\item \textit{History records} provides the history interaction sequence of the target user, which the LLMs can utilize as in-context information for subsequent recommendations~\cite{hou2023large,llara}. 
\item \textit{Reasoning guidance} guides LLMs with essential steps in the reasoning process to make recommendations, incorporating strategies such as chain-of-thought (CoT)~\cite{wei2022chain,hou2023large}. 
\item \textit{Output format} defines the desired format of the LLMs' output in the specific tasks, such as the scores for rating prediction~\cite{geng2022recommendation} and order numbers for ranking tasks~\cite{hou2023large}.
\end{itemize}

\begin{figure}
  \centering
  \includegraphics[width=0.98\textwidth]{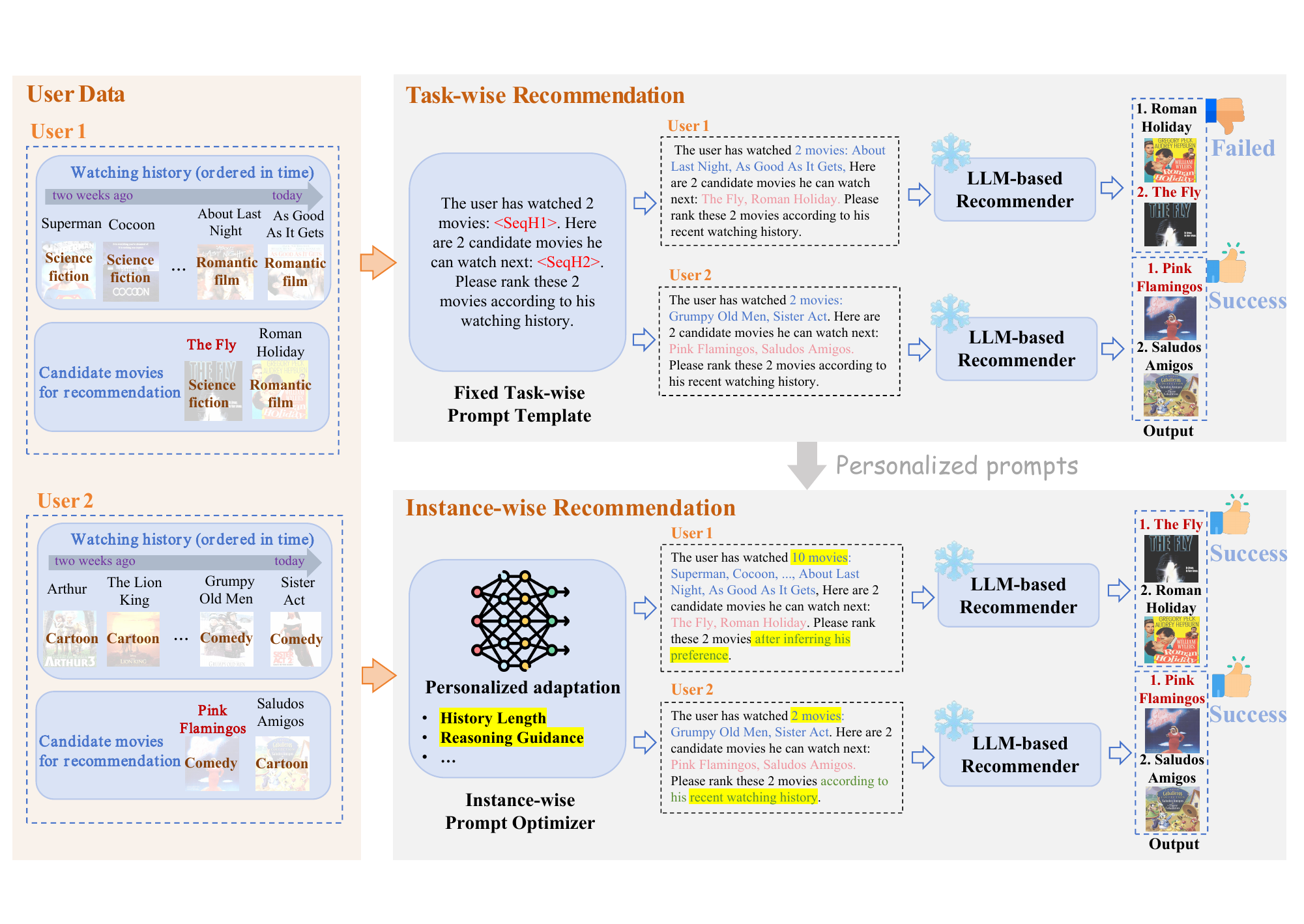}
  \vspace{-8mm}
  \caption{Comparison between task-wise prompting and instance-wise prompting for recommendation. The ground truth for candidate movies that the users will watch next is marked in {\color{red}{red}}. The personalized parts in prompts for different users are highlighted with \colorbox{yellow}{yellow}.}
  \vspace{-6mm}
  \label{fig: methods_comparison}
\end{figure}

However, we argue that applying fixed task-wise prompts for all users~\cite {jin2023instance} may fail to fully explore the capability of LLMs to provide personalized recommendations. First, users' varied intentions pose challenges for the one-fitting-all prompting to analyze dynamically, which may result in mediocre recommendation performance. Take Figure~\ref{fig: methods_comparison} as an example, the first user's preference for the science fiction film ``The Fly'' is reflected in the watching history of 10 movies from two weeks ago. In contrast, the second user's preference for the comedy film ``Pink Flamingos'' is evident in the watching history of the recent two films. 
Thus, task-wise prompts, such as instructing LLMs to recommend based on a fixed length of users' watching history may neglect users' dynamic preferences. Secondly, the performance of LLMs is sensitive to the expression of prompts~\cite{lu2021fantastically}. Fixed prompt templates sacrifice the benefits that varying expressions of prompts can bring to the output of LLMs. Therefore, we propose personalizing \textbf{instance-wise prompts} for different users in the recommendation tasks.

While conceptually appealing, personalizing the discrete prompts for individual users is challenging.  {On the one hand, the methods of manual} creation~\cite{brown2020language} or heuristical crafting~\cite{jiang2020can}) are labor-intensive and resource-expensive to tailor prompts for each user. {On the other hand, methods based on supervised learning \cite{rubin2021learning, liu2022p} are unreliable due to the absence of labels for optimal prompts.} 
To bridge this gap, we propose personalizing prompts automatically with reinforcement learning (RL), drawing inspiration from recent advances in optimizing prompts for natural language processing (NLP) tasks~\cite{deng2022rlprompt,zhang2022tempera}. 
{The advantage of RL \cite{ItemA2C} lies in its ability to learn autonomously through interactions with the environment, rather than depending on manual effort or labeled data. }
However, {existing works which optimize prompts with RL in NLP tasks} either search for tokens within an extensive vocabulary~\cite{deng2022rlprompt} or conduct basic editing operations~\cite{zhang2022tempera} on the prompts, which are constrained by computational resource limitations and the reliability of the generated prompts. 
To apply RL for personalizing instance-wise prompts in the recommendation tasks, two fundamental problems are crucial.

\begin{itemize}[leftmargin=*]
    
    \item How to reduce the search space to enhance efficiency?
    \item How to ensure the quality of the personalized prompts?
\end{itemize}

To tackle these problems, we introduce Reinforced Prompt Personalization (RPP) to personalize high-quality prompts for each user efficiently. Instead of searching word-by-word in the extensive vocabulary \cite{deng2022rlprompt}, we limit the search space by exploring the various expressions of the four patterns at the sentence level, \ie \{\textit{role-playing}, \textit{history records}, \textit{reasoning guidance}, \textit{output format}\}. We view the prompt optimization process as a combinatorial optimization problem with limited search space \cite{anonymous2025apollomilp, liu2024milpstudio, kuang2024rethinking, liu2024deep}. To ensure prompt quality, we meticulously design the diverse expressions of each pattern based on the nature of recommendation tasks, which can guide LLMs to understand users' intent and make recommendations from multiple perspectives. {We specify these high-quality expressions as actions, in contrast to simply treating different editing operations (such as ``add'' or ``delete'') as the actions \cite{zhang2022tempera}. }
The personalized prompts interact with LLMs which function as the environment, to maximize LLMs' recommending performance which serves as the reward. 
Specifically, to select sentences (actions) for the four patterns {with specific expertise} respectively, RPP is designed based on multi-agent RL (MARL) under Centralized Training with Decentralized Execution (CTDE) paradigm ~\cite{lowe2017multi}. Under CTDE, each agent has an individual policy network based on Advantage Actor-Critic ~\cite{peng2018adversarial} algorithm, {as personalizing each specific prompt pattern requires distinct expertise.}
 This global state contains information about the personalized prompts and the ranking results generated by LLMs in the current round. 
 The action space is carefully designed with each action collected through collaborative deliberation between human experts and LLMs. Moreover, to improve the flexibility and quality of the action space, we dynamically refine the selected actions with LLMs during the iterative process to enhance scalability, referred to as RPP+.
 
We assess the effectiveness of our approach in ranking tasks across three widely used public datasets(\ie{MovieLens-1M~\cite{harper2015movielens}, Games~\cite{ni2019justifying}, and Lastfm~\cite{cantador2011second}}). 
Experimental results demonstrate that RPP/RPP+ can significantly improve the recommendation performance of LLMs, outperforming several traditional recommender models (\eg BPRMF~\cite{rendle2012bpr}, LightGCN~\cite{he2020lightgcn}, SASRec~\cite{kang2018self}), few-shot methods (\eg VQ-Rec~\cite{hou2023learning}, UniSRec~\cite{hou2022towards}), and other prompt-based methods (\eg manual prompts~\cite{hou2023large}, enumeration, and GRIPS~\cite{prasad2022grips}). This highlights the significance of tailoring prompts for each individual user and the superiority of RPP/RPP+ in enhancing the recommendation ability of LLMs.
Our key contributions are as follows:
\begin{itemize}[leftmargin=*]
\item 
We propose Reinforced Prompt Personalization (RPP/RPP+), a general prompt optimization framework to tailor instance-wise prompts for individual users.
\item
Our methodology generates personalized prompts at the sentence level with MARL, achieving the balance between search efficiency and prompt quality.
\item 
Extensive experiments on three benchmark datasets demonstrate the superiority of RPP/RPP+, highlighting the necessity of personalizing instance-wise prompts.
\end{itemize}

%% file: 3_problem_formulation.tex
\section{Preliminaries}
Here, we introduce task-wise prompting and instance-wise prompting in the recommendation tasks.

% \subsection{Task-wise Prompting and Instance-wise Prompting}
% \begin{wrapfigure}{r}{0.5\textwidth}
% \vspace{-8mm}
% \footnotesize
% \begin{tcolorbox}[colback=black!4!white,colframe=black!75!black,title=Task-wise Prompt Example,top=2pt, bottom=2pt]
% \textbf{Persona:}\\
% You are a movie expert.\\
% \textbf{Records:}\\
% I've watched these movies {\color{red}{$\langle$SeqH1$\rangle$}} recently.\\
% \textbf{Recipe:}\\
% Please rank these candidate movies {\color{red}{$\langle$SeqH2$\rangle$}} after inferring my preference from my watching history.\\
% \textbf{Template:}\\ Please only output the ranking results with order numbers. Do not explain the reason or include any other words.''
% \end{tcolorbox}
% \noindent  
% \end{wrapfigure}
% Most LLM-based recommender systems frame recommendation as a language modeling task, using discrete prompt templates defined at the task level, with a fixed prompt format shared across all users regardless of the user's personalized information.
% The fixed prompt format typically consists of ``persona'', ``records'', ``recipe'', and ``template'' patterns. 
% Here is an example of task-wise prompt format, where {\color{red}{$\langle$SeqH1$\rangle$}} and {\color{red}{$\langle$SeqH2$\rangle$}} are the lists of movie titles from the watching history and candidate movies, respectively.

\subsection{Task-wise Prompting}
Most LLM-based recommender systems frame recommendation as a language modeling task~\cite{wang2023zero,gao2023chat,dai2023uncovering}, with a fixed prompt template shared across all users in a task. 
The fixed task-wise prompts typically consist of four patterns, including ``role-playing'', ``history records'', ``reasoning guidance'', and ``output format'', as exemplified as follows:
\begin{center}
\begin{tcolorbox}[colback=black!3!white,colframe=black!75!black,title=Task-wise Prompt Example,width=0.8\textwidth,top=2pt, bottom=2pt]
\textbf{Role-playing:}\\
You are a movie expert.\\
\textbf{History records:}\\
I've watched these movies {\color{red}{$\langle$SeqH1$\rangle$}} recently.\\
\textbf{Reasoning guidance:}\\
Please rank these candidate movies {\color{red}{$\langle$SeqH2$\rangle$}} after inferring my preference from my watching history.\\
\textbf{Output format:}\\ Please only output the ranking results with order numbers. Do not explain the reason or include any other words.''
\end{tcolorbox}
\end{center}
\noindent where {\color{red}{$\langle$SeqH1$\rangle$}} and {\color{red}{$\langle$SeqH2$\rangle$}} are the lists of movie titles from the watching history and candidate movies, respectively.
Each sentence in the prompt template correspondings to a fixed pattern that remains unchanged regardless of the difference of users. The one-size-fits-all approach lacks the flexibility to adapt the prompt to catch individual users’ preferences, leading to inaccurate recommendation results.

\subsection{Instance-wise Prompting}
To match the need for personalization, going beyond task-wise prompting, we introduce instance-wise prompting for recommendation, which tailors prompts for each user instance to elicit high-quality recommendations.
% aims to optimize personalized prompts for each user instance, so as to obtain high-quality recommendations from LLMs.
Formally, for a user $u$, we take his/her historical interactions $\mathcal{H} = \{i_k\}_{k=1}^n$ paired with the candidate items $\mathcal{C} = \{i_j\}_{j=1}^m$ as the input $x$. 
Then, the objective of prompt personalization can be formulated as:
\begin{gather}
p\ast=\mathrm{argmax}_{p\in \mathbb{P}}  \mathrm{R} (y_{LM}(p, x)),
\end{gather}
where $\mathbb{P}$ is the space of possible prompts. The LLM observes $x$ embedded in the prompt to yield recommendation results $y_{LM}$, reflecting the user preference on candidate items. The function $\mathrm{R}(\cdot)$ quantifies the alignment between the predicted results and ground truth, with higher values indicating greater alignment. Thus, the task becomes identifying $p\ast$, the optimal prompt that maximizes alignment for each individual user.

%% file: 4_methodology.tex
\section{Methodology}

\begin{figure*}[t]
    \centering
    \vspace{-4mm}
    \includegraphics[width=1\textwidth]{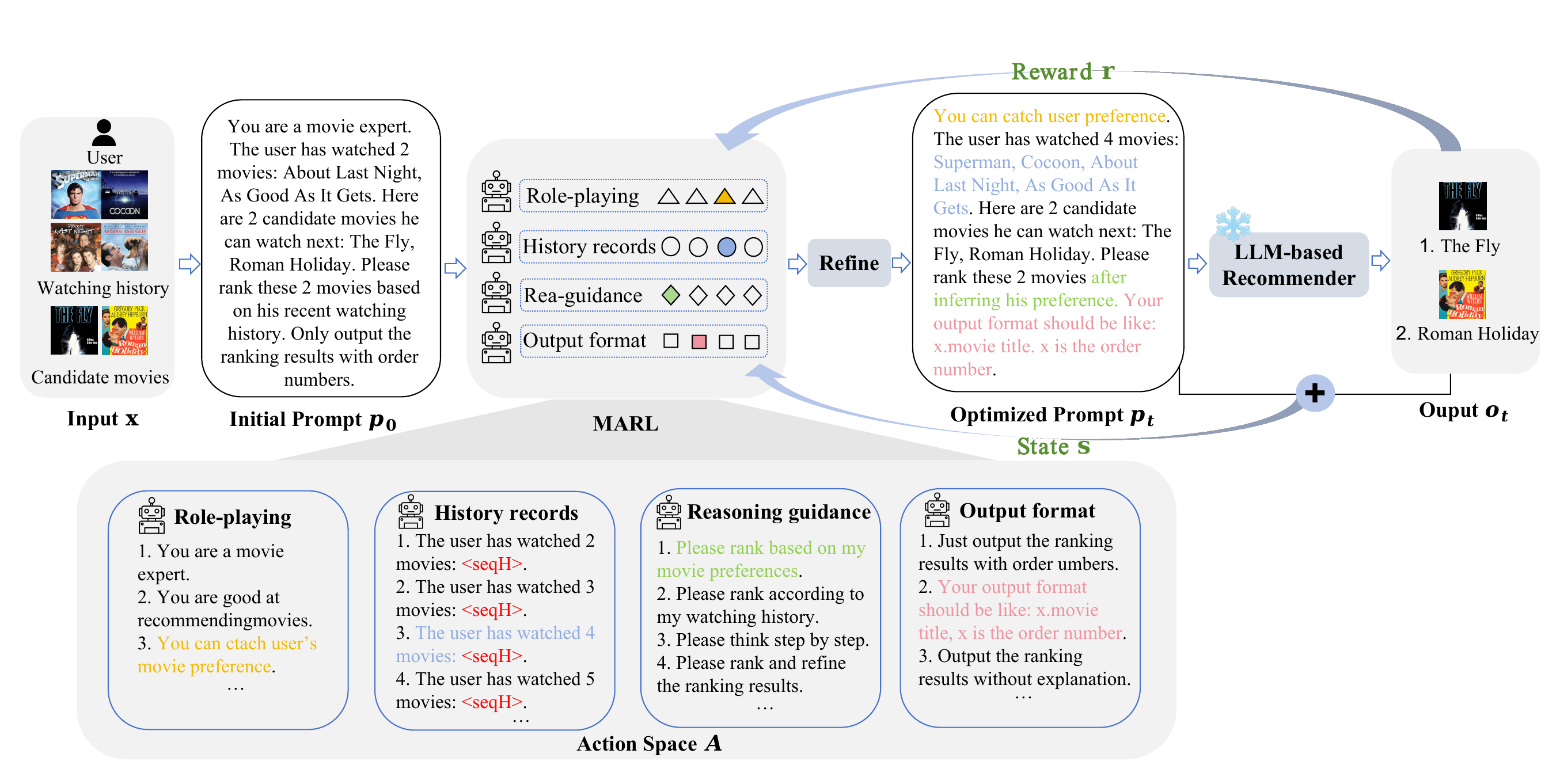}
     % \vspace{-8mm}
    \caption{The framework of our proposed RPP/RPP+. {MARL serves as the core component to personalize instance-wise prompts with four distinct patterns, using corresponding agents. It is trained iteratively to maximize rewards based on the outputs of the frozen LLM-based recommender. Once trained, MARL can select optimal actions from four patterns to generate personalized prompts for each user based on their data, effectively prompting the LLM-based recommender for tailored recommendations. In addition to RPP, the ``Refine'' block is designed for RPP+ to enhance the flexibility and quality of the selected actions, utilizing other LLMs to dynamically refine the selected actions before prompting the LLM-based recommender.}}
    \label{fig:framework}
 \vspace{-4mm}
\end{figure*}

To enhance LLMs' recommendation capability with effective prompts, we introduce RPP/RPP+ to personalize instance-wise prompts with MARL, where RPP+ enhances RPP by incorporating a ``refine'' block. {Comprising four actor-critic networks, the MARL serves as the core component to select actions from the four patterns and create personalized instance-wise prompts for different users. These optimized prompts are then used to prompt an LLM-based recommender for recommendations, while the MARL is trained iteratively to maximize rewards from the outputs of LLM, as shown in Figure \ref{fig:framework}.} In this section, we begin by formulating prompt personalization as a Markov Decision Process, followed by a detailed description of RPP/RPP+'s components, including the action space, state space, reward function, and policy architecture.
 
 \subsection{Formulation of RL for Prompt Personalization}
Manual creation or heuristical crafting may fail to personalize discrete prompts automatically and efficiently.
Building upon the idea of using reinforcement learning for prompt personalization, we view the prompt generation as selecting actions from natural language space and formulate the problem as a Markov Decision Process (MDP) $\langle \mathcal{S}, \mathcal{A}, T, R, \gamma \rangle$ to optimize prompts in an instance manner, wherein,
\begin{itemize}[leftmargin=*]
    \item $\mathcal{S}$ is the state set, where $\Mat{s} \in \mathcal{S}$ indicates the state containing information about the current prompt and recommendation result. 
    
    \item $\mathcal{A}$ is the action set, where each action $a \in \mathcal{A}$ corresponds to a sentence selected from a specific pattern (\eg {``You are a movie expert.'' for ``role-playing''}).
    
    \item $T$ is the transition function, where $t(\Mat{s}_{t+1}|\Mat{s}_t, a_t)$ is the transition possibility that action $a_t$ in state $\Mat{s}_t$ at time $t$ will lead to state $\Mat{s}_{t+1}$ at time $t+1$. In our task, $t(\Mat{s}_{t+1}|\Mat{s}_t, a_t)$ represents the possibility of LLMs' recommendation results at the $t+1$ iteration.
    
    \item $R$ is the reward function,  $r(\Mat{s}_t, a_t)$ determines the reward $r \in R$ for agent after interacting with the LLMs. To integrate with recommendation tasks, we define the reward $r$ as the recommendation performance evaluated according to LLMs' answers.
    
    \item $\gamma$ is the discount factor that controls the agent's emphasis on future rewards compared to immediate rewards.
\end{itemize}

For a user $u$, we initialize prompt $p_0$ with sentences randomly selected from $K$ patterns (\ie{role-playing, history records, reasoning guidance, and output format}).
As depicted in Figure~\ref{fig:framework}, we employ a multi-agent RL (MARL) under the Centralized Training with Decentralized Execution (CTDE) paradigm~\cite{lowe2017multi}. Each agent $z_k$ is responsible for personalizing a specific pattern with individual action space $\mathcal{A}_k$, where $k\in\{1, 2, ..., K\}$, $K$ is the number of agents (or patterns). 
Prompt personalization can be formulated as seeking policies $\pi_k(a_t^{(k)}|\Mat{s}_t)$ that selects actions $a_t^{(k)}$ from the action
set $\mathcal{A}_k$ based on the global state $\Mat{s}_t$, generating a prompt $p_t=concat(a_t^{(1)},\ldots, a_t^{(K)})$ at step $t$. The generated prompt $p_t$ interacts with LLMs (environments) to enhance LLMs' recommendation performance (reward) by maximizing the objective of our multi-agent.  
Formally, the objective can be defined as:
\begin{equation}
\begin{aligned}
Q\left(\Mat{s}_t,\{a_t^{(k)}\}_{k=1}^K\right)&={\mathbb{E}}_{
\underset{\{a_{t'}^{(k)} \sim \pi_k \}_{k=1}^K}{\Mat{s}_{t'}\sim \mathcal{S}}}\left[\sum_{t'=t}^{\infty}\gamma^{t'-t}\cdot r\left(\Mat{s}_{t'},\{a_{t'}^{(k)}\}_{k=1}^K \right) \right],\\
\max \limits_{ \{\pi_k\}_{k=1}^K}\mathcal{J}\left(\{\pi_k\}_{k=1}^K\right)&={\mathbb{E}}_{
\underset{\{a_{t}^{(k)} \sim \pi_k \}_{k=1}^K}{\Mat{s}_{t}\sim \mathcal{S}}}\left[Q(\Mat{s}_t,\{a_t^{(k)}\}_{k=1}^K)\right],
\end{aligned}
\end{equation}
 where $Q\left(\Mat{s}_t,\{a_t^{(k)}\}_{k=1}^K\right)$ denotes the expected cumulative reward for MARL at step $t$ which can be estimated by Q-value function. By optimizing the policies $\pi_k$ and maximizing the objective $\mathcal{J}$, we can obtain the optimal prompt $p_t$ for each user during iterations.

\subsection{Action Space}
Designing an effective action set is crucial when optimizing prompts with reinforcement learning. The key challenge is balancing search efficiency and prompt quality. 
To manage this trade-off, we develop the action space comprising several subspaces $\{\mathcal{A}_k\}_{k=1}^K$, each representing a distinct pattern and equipped with multiple sentences as actions. Inspired by previous studies~\cite{shanahan2023role,hou2023large,zhang2023recommendation} that have investigated task-wise prompts utilizing role-playing, history records, reasoning guidance, and output format patterns, we explore personalizing instance-wise prompts by optimizing these four patterns. To improve searching efficiency, we optimize each pattern by selecting actions (sentences) from $\{\mathcal{A}_k\}_{k=1}^K$, as depicted in Figure~\ref{fig:framework}. The sentence-level optimization for these four patterns enhances the efficiency of RPP/RPP+, rather than optimizing word-by-word within the vast vocabulary. 
Moving to the goal of ensuring prompt quality, each pattern $\mathcal{A}_k$ consists of multiple sentences meticulously designed for the recommendation tasks, free from ambiguous words, and encompassing diverse perspectives for LLMs to analyze. These sentences represent the possible actions in action space $\mathcal{A}_k$ for the corresponding agent $z_k$.

For implementation, the action set is jointly established by both humans and LLMs for specific tasks, such as movie ranking tasks. Each action is meticulously crafted by the LLMs and rigorously screened by humans to ensure its task relevance and encompass diverse analytical perspectives, thereby guaranteeing prompt quality. Moreover, to enhance the flexibility of personalized prompts, we leverage LLMs to refine the actions selected at each iteration, enabling a more adaptable expression for the generated prompts. And we name it as RPP+.
Below we explain the design of the action subset for each pattern in detail.

\smallskip \noindent  \textbf{Role-playing: } This pattern assigns a specific role to LLMs~\cite{shanahan2023role,shao2023character} to generate responses that are consistent with a particular profession. For instance, embodying the role of LLM as a movie expert allows LLMs to leverage their extensive knowledge about the movie domain.
In the context of recommendation, the actions of the ``role-playing'' pattern include:
% For the recommendation task, the sentences (actions) for the ``persona'' pattern can be classified into three categories.
1) Positioning the LLM as an expert to leverage the domain knowledge (\eg ``You are a movie expert.'')
2) enhancing recommendation capabilities of LLM by assigning them as a recommender  (\eg ``You are good at recommending movies.'')
3) instructing LLM to discern users' inner intents, akin to a psychologist's role, such as ``You are good at catching people's movie interest.''.

\smallskip \noindent \textbf{History records: }
The user's historical interactions constitute critical contextual information for eliciting high-quality recommendation~\cite{liu2023chatgpt,hou2023large}. Intuitively, certain users exhibit short-term interests best captured by recent interactions, while others having long-term preferences require a longer interaction history.
Therefore, we personalize history records by varying the interaction sequence length to match users' dynamic preferences.
Specifically, with an initial length $l_0$, at each step $t$, the agent's action is deciding the number of additional interactions $\hat{l}_t$ to introduce in chronological order, that is, providing $l_{t} = l_{t-1} + \hat{l}_t$ historical context from near to far in time. By formulating the sequence expansion as action, the agent can learn an adaptive policy to tail each user's sequence to the short versus long-term interests.

\smallskip \noindent  \textbf{Reasoning guidance: }
This pattern offers guidelines for LLMs to systematically decompose complex tasks into multiple reasoning steps, incorporating strategies such as chain-of-thought (CoT)~\cite{wei2022chain}, reflexion~\cite{shinn2023reflexion}, and refinement~\cite{madaan2023self}. 
Since recommendation lies in inferring user intentions based on their behaviors, it is crucial to introduce intermediate steps to move from the surface to the essence.
We include six action sub-categories within the ``reasoning guidance'' pattern:
1) direct recommendation without intermediate steps (\eg ``Please rank these candidate movies in order of priority from highest to lowest.'') 2) explicitly instructing LLM to recommend based on the user's interaction history (\eg ``Please rank these candidate movies according to my watching history.'')
3) prompting LLM to infer user preference before suggesting items (\eg ``Please rank these candidate movies based on my movie preferences which are inferred from my watching history.'') 
4) allowing LLM to deduce without specific instructions (\eg ``Please rank these candidate movies and think step by step.'') 
5) utilizing the refinement and update mechanism of LLMs in the process of generating recommendation results (\eg ``Please rank these candidate movies and refine the ranking results according to my watching history.'')
6) requiring LLM to calculate the similarity between candidate items and user preference before recommending items (\eg ``Please rank these candidate movies after calculating the similarity between them and my movie preferences, according to my watching history.'').

 \smallskip \noindent  \textbf{Output format: }
This pattern defines the desired format of the LLMs' output in recommendation tasks, thereby avoiding invalid output and facilitating accurate evaluation. 
We provide multiple output format options tailored to the recommendation task:
1) outputting titles and order numbers, with each item on a separate line (\eg ``Please only output the ranking results with order numbers. Split these order
numbers with a line break '')
2) outputting the recommendation results with order numbers without any unnecessary steps in the answer (\eg ``Attention! Just output the ranking results with order numbers and ignore any unnecessary steps.'') 
3) answering without explanatory text (\eg ``Please only output the ranking results with order numbers. Do not explain the reason or include any other words '') 
4) exemplifying the output format~\cite{brown2020language,zhao2021calibrate} (\eg ``Please
only output the ranking results with order numbers. Your output
format should be like this: x. movie title, x is the order number'').

We should emphasize that the presented action set serves as a representative example of personalized prompt optimization in movie recommendations. However, this framework can be generalized to various recommendation tasks (\eg rating prediction, item ranking) and domains (\eg games, music) through the substitution of task requirements or domain-specific terminology. Moreover, the current action set can be expanded with additional choices belonging to each pattern. These demonstrate the flexibility and scalability of the proposed framework. 

\smallskip \noindent {\textbf{Refine actions in RPP+: }In addition to RPP, the ``refine'' block designed for RPP+ can dynamically refine the selected actions with other LLMs before prompting the LLM-based recommenders during each iteration, which can automatically enhance the flexibility and quality of the selected actions. To refine the selected sentence action with LLMs, we use the prompt as ``Please refine this sentence to effectively prompt LLMs for recommendations''. After the ``refine'' block of RPP+, we obtain a more polished sentence for prompting compared to that generated by RPP.}

\subsection{State Space}
The state space provides key observations for informing the agent's actions. An effective state should include sufficient information about the current environment. 
Toward this goal, we initialize the state $\Mat{s}_0$ with user-specific information. Subsequently, at step $t$, we update state $\Mat{s}_t$ with the current prompt $p_t$ and LLMs' ranking output $o_t$ to include sufficient information about the environment. 
We now detail the process of acquiring these states.

\smallskip \noindent 
\textbf{Initializing state $\Mat{s}_0$:} 
% In our framework, differentiating prompts between users is a critical step.
{Since state information after decisions is unavailable initially, we initialize the state $\Mat{s}_0$ by incorporating users' personalized features to differentiate between them.} Here we set the user-specific features as user embedding derived from the traditional recommender models (\eg LightGCN~\cite{he2020lightgcn}), which can capture collaborative signals with graphs \cite{WWW_GIF,FCS_wu}. Formally, given the ID information of user $u$ and item $i$, we obtain their embeddings:
\begin{equation}
\begin{array}{c}
\Mat{u}, \Mat{i}=f(u, i), \\
\Mat{s}_0=\Mat{u},
\label{eq: S_0}
\end{array}
\end{equation}
where $\Mat{u}$ and $\Mat{i}$ denote the representation of user $u$ and item $i$, respectively, $f(\cdot)$ represents the encoder of traditional recommender models. Based on the personalized information of each user, RPP/RPP+ makes policy to obtain the initial actions $a_0^{(k)}$ and initial prompts $p_0$.

\smallskip \noindent 
\textbf{Updating state $\Mat{s}_t$: }
To assess the prompt quality and LLMs' output comprehensively, we view the current prompt $p_t$ and LLMs' ranking output $o_t=[\hat{i}_1, \cdots, \hat{i}_j, \cdots, \hat{i}_M]$ from LLMs as the shared state of agents, where $\hat{i}_j$ is the $j$-th recommended item in the output list, $M$ is the number of candidates.
To obtain the representation of states at step $t$, we leverage a pre-trained language model (\ie BERT~\cite{devlin2018bert}) and a sequential model (\ie GRU) to encode $p_t$ and $o_t$, respectively.
Formally, we have,
\begin{equation}
\begin{aligned}
    \Mat{e}_{t}^{(p)}&=BERT(p_t),\\
    \Mat{e}_{t}^{(o)}&=GRU(\Mat{\hat{i}}_1,
    \Mat{\hat{i}}_2, \cdots, \Mat{\hat{i}}_M), \\
    \Mat{s}_t&=\Mat{e}_{t}^{(p)}+\Mat{e}_{t}^{(o)},
\label{eq: S_t}
\end{aligned}
\end{equation}
\noindent where $\Mat{\hat{i}}_j$ denotes the embedding of item $\hat{i}_j$, $\Mat{e}_{t}^{(p)}$ and $\Mat{e}_{t}^{(o)}$ are the representations of prompt $p_t$ and ranking result $o_t$, respectively.

\subsection{Actor-Critic Based Architecture of the Multi-agent and Reward Function}
To personalize each pattern individually with the corresponding agent, the Multi-Agent Reinforcement Learning (MARL) is structured with four A2C agents~\cite{peng2018adversarial} under CTDE \cite{lowe2017multi,sunehag2017value}, where each agent has independent parameters but has access to the global state, as shown in Figure~\ref{fig:framework}. 
The Actor and Critic of each agent are two-layer fully connected neural networks parameterized by $\theta_g^{(k)}$ and $\theta_h^{(k)}$, denoted as $g^{(k)}(\cdot)$ and $h^{(k)}(\cdot)$, respectively. Specifically, each Actor $g^{(k)}(\cdot)$ makes policy to obtain action $a_t^{(k)}$ and its corresponding probability $prob_t^{(k)}$ based on the state $\Mat{s}_t$, while each Critic $h^{(k)}(\cdot)$ evaluates the value $v_t^{(k)}$ of state $\Mat{s}_t$. Formally, we have:
\begin{equation}
    \begin{aligned}
       a_t^{(k)}, prob_t^{(k)}=g^{(k)}(\Mat{s}_t), \\
       v_t^{(k)}=h^{(k)}(\Mat{s}_t). 
    \end{aligned}
\end{equation}

Since the aim is to boost the recommendation ability of LLMs by personalizing instance-wise prompts, we leverage Normalized Discounted Cumulative Gain (NDCG) directly as the reward, which is the performance metric for ranking tasks. Here we adopt $NDCG@M$ as the reward evaluated according to the answer of LLMs, where $M$ is the number of candidate items {and we set it as 10 in our experiments}. Formally, given the recommendation output $o_t$ at step $t$, we can define the reward $r_t$ as below:
\begin{equation}
    \begin{aligned}
    r_t=\mathrm{NDCG@M}(o_t)
    \label{eq: reward}
        \end{aligned}
\end{equation}
Given the current and future reward $r_t, r_{t+1},\cdots, r_{t+n}$, current and future value $v_t^{(k)}, \cdots, v_{t+n}^{(k)}$ evaluated by the $k^{th}$ Critic, we can define the optimization objectives $L_a^{(k)}$ and $L_c^{(k)}$ for Actors and Critics as follows:
\begin{align}
       \hat{R}_t=r_{t+1}+\gamma r_{t+2}+\cdots+\gamma^{n-1}r_{t+n}+\gamma^n v_{t+n}^{(k)},
\end{align}
\begin{align}
       L_c^{(k)}&=\frac{1}{N}\sum_t ({\hat{R}_{t-1}}-v_t^{(k)}), \\ L_a^{(k)}&=\frac{1}{N}\sum_t\mathrm{log}(prob_t^{(k)}({\hat{R}_{t-1}}-v_t^{(k)})),
       \label{eq: loss}
\end{align}
where $\gamma$ is a discount factor for emphasis on short-term reward while assigning less weight to long-term reward, $\hat{R}_t$ denotes the cumulative sum of future reward.

\subsection{Algorithm}
The algorithm for our framework is presented {in Algorithm \ref{alg1}}.
% \vspace{-2mm}
\begin{algorithm}
\caption{Training Reinforced Prompt Personalization for each epoch.}
\label{alg1}

\begin{algorithmic}[1]

\REQUIRE Initial Actors $\{g^{(k)}\}_{k=1}^4$ and Critics $\{h^{(k)}\}_{k=1}^4$, initial state $\Mat{s}_0$, learning rate $\alpha_a, \alpha_c$ for Actors and Critics
\ENSURE Actor networks $\{g^{(k)}\}_{k=1}^4$

\FOR{t={0, 1, ...}}
    
         \FOR{k=1:4}
            \STATE 
            $a_t^{(k)},prob_t^{(k)}\gets g^{(k)}(\Mat{s}_t)$
            \STATE
            $v_t^{(k)} \gets h^{(k)}(\Mat{s}_t)$
        \ENDFOR
        \STATE $p_t=refine(concat(a_t^{(1)}, \ldots, a_t^{(K)}))$ 
        % \STATE $p_t=refine(p_t)$ \hfill {\color{gray} \# the refine module of RPP+}
         \STATE $o_t\gets y_{LM}(p_t,x)$ \hfill {\color{gray} \#interact with LLMs}
        \STATE Update the state $\Mat{s}_{t+1}$ according to Equation~\ref{eq: S_t}
        \STATE Obtain the reward $r_t$ according to Equation~\ref{eq: reward}
        \FOR{k=1:4}
        \STATE $\theta_g^{(k)} \gets \theta_g^{(k)}-\triangledown_{\theta_g^{(k)}} 
        L_a^{(k)} $\hfill {\color{gray} \# optimize Actors' parameters}
        \STATE $\theta_h^{(k)} \gets \theta_h^{(k)}-\triangledown_{\theta_h^{(k)}} L_c^{(k)} $\hfill {\color{gray} \# optimize Critics' parameters}
        \ENDFOR
        
        % \STATE Optimize the parameters $\theta_i^a$ and $\theta_i^c$ of the multi-agent according to the Equation~\ref{eq: loss}
  
\ENDFOR

\end{algorithmic}

\end{algorithm}
% \vspace{-4mm}

%% file: 5_experiments.tex
\section{Experiments}
In this section, we conduct extensive experiments with three public datasets in RSs to evaluate our proposed framework RPP/RPP+ on the ranking tasks and answer the following questions:
\begin{itemize}[leftmargin=*]

\item  RQ1: Can personalizing instance-wise prompts with our framework RPP/RPP+ effectively enhance the recommendation ability of LLMs?

 \item RQ2: How well can the framework RPP/RPP+ generalize on diverse types of LLMs?
 
 \item RQ3: What's the respective contribution that each pattern of the framework can make to our task?
 
 \item RQ4: To what extent is the framework sensitive to variations in hyper-parameters?
\end{itemize}
We evaluate the effectiveness of RPP/RPP+ on LLaMa2-7B-chat, comparing it against various baselines including traditional recommender models, few-shot methods, and prompt-based methods. Besides, we extend the evaluation on diverse frozen pre-trained LLMs to measure the RPP/RPP+'s generalization capability. To gain insights into the impact of different patterns in prompts, we conduct an ablation study on personalizing each pattern. Furthermore, we investigate the sensitivity of RPP/RPP+ to hyper-parameters. The case study is employed to provide a tangible illustration of the significance of personalizing prompts in enhancing LLMs' recommendation ability.

\subsection{Experimental Settings}

\subsubsection{Datasets}
We conduct experiments on three public datasets for RSs --- MovieLens-1M (ML-1M)~\cite{harper2015movielens}, Games~\cite{ni2019justifying}, and Lastfm~\cite{cantador2011second}. ML-1M is a widely-used movie dataset that contains 1 million ratings given by users on movies. The Games dataset is a collection of user reviews and ratings for various video games available on the Amazon platform. The Lastfm dataset contains a rich collection of user listening histories, including user profiles, artist names, and track names.
Following previous work~\cite{hou2023large}, the reviews or ratings are regarded as interactions. The users and items with fewer than five interactions are excluded from the analysis. The interactions for each user are organized sequentially based on their timestamps, arranging the oldest interactions first. The detailed statistics of the datasets are
listed in Table~\ref{table: dataset}.
\begin{table}[htb]
\centering
\renewcommand\arraystretch{1.2}
% \vspace{-2mm}
\caption{Statistics of datasets.}
% \vspace{-2mm}
\label{table: dataset}
\setlength{\tabcolsep}{1.5mm}{\begin{tabular}{lccc}
\toprule
\textbf{Dataset} & \#\textbf{Users} & \#\textbf{Items} & \#\textbf{Interactions} \\
\toprule
ML-1M&6,040&3,885  &1,000,210\\
Games& 50,545 & 16,858&389,718\\
Lastfm&2,100&17,632&92,835\\
\bottomrule
% \vspace{-6mm}
\end{tabular}}
\end{table}
% \vspace{-4mm}

\subsubsection{Baselines}
We employ several baseline models for comparison, categorizing them into three types: (1) traditional recommender models, including Pop,  BPRMF~\cite{rendle2012bpr}, LightGCN~\cite{he2020lightgcn}, and SASRec~\cite{kang2018self};
(2) few-shot models, such as BM25~\cite{robertson2009probabilistic}, VQ-Rec~\cite{hou2023learning} and UniSRec~\cite{hou2022towards}; (3) prompt optimization methods, including manual prompts~\cite{petroni2019language,schick2020exploiting}, enumeration that is a kind of heuristic prompts~\cite{jiang2020can,gao2020making,prasad2022grips}, and GRIPS~\cite{prasad2022grips}. 

\smallskip \noindent 
\textbf{Traditional recommender models: }
Pop relies on item popularity for recommendations but overlooks personalization. BPRMF~\cite{rendle2012bpr}
combines Bayesian methods and matrix factorization techniques
to decompose the user-item interaction matrix into a latent factor
matrices. LightGCN~\cite{he2020lightgcn} is a simplified and efficient framework designed for recommendation tasks using graph convolutional networks (GCNs). SASRec~\cite{kang2018self} leverages a self-attention encoder to comprehensively analyze the long-term dependencies present in user behaviors. And it predicts subsequent interactions through a feed-forward network. They are evaluated on the candidate sets after
training on the training dataset with all interactions following the setting of LLMRank~\cite{hou2023large}

\smallskip \noindent
\textbf{Few-shot methods:} 
BM25~\cite{robertson2009probabilistic} is a versatile information retrieval algorithm that enhances the accuracy of item ranking by considering item frequency. VQ-Rec~\cite{hou2023learning} suggests an item representation scheme to acquire Vector-Quantized item representations specifically designed to facilitate transferable sequential recommendation;
UniSRec~\cite{hou2022towards} leverages item description text to acquire transferable representations, which can be implemented by a lightweight item encoding architecture and two contrastive pre-trained tasks. Following the setting of LLMRank~\cite{hou2023large}, we leverage their publicly available pre-trained models to evaluate their performance on ranking the candidate items without training on the target dataset.

\smallskip \noindent 
\textbf{Prompt-based methods: } 
Manual prompts~\cite{petroni2019language,schick2020exploiting} are prompts designed by humans to guide LLMs' behavior, relying on human expertise and domain knowledge. Here we meticulously craft a task-wise prompt for the ranking tasks, encompassing the key patterns within a prompt. Enumeration entails automatically searching within several possible prompts and selecting the best one. Here we design diverse prompts with variations for the ranking tasks, adopting enumeration followed by selection. GRIPS~\cite{prasad2022grips} performs edit operations
on prompts automatically for LLMs with gradient-free search. Here we adapt it as an instance-wise prompting baseline for ranking tasks.

\subsubsection{Frozen Pre-trained LLMs}
To evaluate the generalization ability of RPP/RPP+, we extend the experiments on diverse frozen pre-trained
LLMs, including LLaMa2, ChatGPT, and Alpaca. They
serve as the representative LLMs of three types respectively: the
open-source LLM, which can be inferred using pre-trained parameters; the black-box LLM, which solely offers an accessible API; and the fine-tuned LLM, tailored specifically for the recommendation task. LLaMa2 is an open-source LLM released by Meta whose performance has been widely recognized. Here we select the version of LLaMa2-7B-chat as the frozen LLM. ChatGPT is a black-box LLM, which is excellent across a wide array of tasks. We conduct experiments by utilizing the API of GPT-3.5-turbo from OpenAI and set the temperature to 0.2. Alpaca is a lightweight
instruction-tuning strategy based on LLaMa. We select LLama2-
7B as the backbone and fine-tune Alpaca with Lora~\cite{hu2021lora}, which
adds trainable rank decomposition matrices to each layer of the
transformer architecture in the frozen LLM~\cite{bao2023tallrec}, enabling easier and more efficient fine-tuning. We transform the recommendation data for ranking tasks into instruction-tuning data, ensuring that LLMs grasp the specific nuances of the ranking task. The format of the
data for instruction tuning can be articulated as follows:

\begin{center}
\begin{tcolorbox}[colback=black!4!white,colframe=black!75!black,title=Data Example for Instruction Tuning,width=0.8\textwidth,top=4pt, bottom=4pt]
\textbf{Instruction}: ``Here is my interaction history: {\color{red}{$\langle$SeqH1$\rangle$}}. There are several candidate items: {\color{red}{$\langle$SeqH2$\rangle$}}. Please rank these candidate items by measuring the possibilities that I would like to interact with. Please think step by step. Merely output the ranking results and order numbers. Split these order numbers with line break.''\\

\textbf{Output}: \color{red}{$\langle$SeqH3$\rangle$}
\end{tcolorbox}
\end{center}

% \smallskip \noindent
% \textbf{Instruction}: \texttt{``Here is my interaction history: {\color{red}{$\langle$SeqH1$\rangle$}}. There are several candidate items: {\color{red}{$\langle$SeqH2$\rangle$}}. Please rank these candidate items by measuring the possibilities that I would like to interact with. Please think step by step. Merely output the ranking results and order numbers. Split these order numbers with line break.''}

% \smallskip \noindent 
% \textbf{Output}: \texttt{\color{red}{$\langle$SeqH3$\rangle$}}

\smallskip 
\noindent where {\color{red}{$\langle$SeqH1$\rangle$}}, {\color{red}{$\langle$SeqH2$\rangle$}} and \texttt{{\color{red}{$\langle$SeqH3$\rangle$}}} are the lists of item titles. Subsequently, we employ the formatted "Instruction" and "Output" as the instruction examples to train Lora on ranking tasks. Then we view the fine-tuned Alpaca as a frozen LLM to conduct experiments.

\subsection{Implementation Details}
We configure the initial length of interaction history $l_0$ as 1 and the number of candidate items to be $M=10$. In our experiments, we set 3, 9, and 5 sentence choices for ``role-playing'', ``reasoning guidance'', and ``output format'' patterns in the action spaces respectively. Candidate items are randomly selected from the training dataset and include a ground-truth item sourced from the test dataset. Therefore, for ranking tasks, the HitRatio@10 metric is always 1, rendering this metric meaningless. We set the $\gamma$ in the reward as 0.95. To reduce the randomness of LLMs, we set the temperature as 0.2, since a lower temperature favors more conservative and deterministic outputs. Besides, we report the average performance with corresponding standard deviations of prompt-based methods and RPP/RPP+, based on at least five repeated runs. Traditional recommender models and few-shot-based methods are executed on an NVIDIA RTX 3090Ti, while prompt-based methods and our framework, based on LLMs, run on an NVIDIA A100. During the training phase, we randomly select 200 users as training examples to optimize the multi-agent strategy employed in RPP/RPP+. We terminate the iterative training process if the highest NDCG@10 does not update over 7 iterations or if the total iterations exceed 15. Subsequently, during the testing phase, the multi-agent, now equipped with an optimized policy, undergoes evaluation on a separate set of 100/200 test examples. {Our experiments show that on the training set, after three rounds of prompt optimization, the optimized prompts can achieve superior performance on ChatGPT for most instances, while it requires two, seven, and eight iterations on the ML-1M, Games, and Lastfim datasets for LLaMa2-7B-chat, respectively, and approximately eight iterations for Alpaca. Consequently, during the inference process, we terminate the iterations at the corresponding rounds for different datasets and LLMs.} We only require users' collaborative embeddings to initialize the state, along with their interaction history and candidate items as input. {To address invalid responses from LLMs, we carefully design prompts to enforce LLMs' output format, reducing these occurrences. Additionally, we trim and pad the LLMs' recommendations as 10 while processing invalid responses.}

\begin{table*}[htb]
\renewcommand\arraystretch{1.2}
 % \vspace{-4mm}
    \caption{Performance comparison of our framework on LLaMa2-7B-chat with baselines across diverse datasets. The results of the best-performing baseline are denoted with an \underline{underline}. The results of our method which outperform other prompt-based methods are indicated with $\uparrow$, while those that surpass the best-performing baseline are highlighted in {\color{red}{red}}. The performance improvement of RPP/RPP+ compared with the best-performing task-wise prompting method (\ie Enumeration) is listed in the ``improvement''.}
 % \vspace{-3mm}
\label{table:main_all}
\centering
\small
\setlength{\tabcolsep}{1.2mm}
{\begin{tabular}{llcccccccc}
\toprule
\multirow{2}*{methods} & \multicolumn{3}{c}{ML-1M} & \multicolumn{3}{c}{Games}&\multicolumn{3}{c}{Lastfm}  \\
\cmidrule(lr){2-4}\cmidrule(lr){5-7}\cmidrule(lr){8-10}
 &N@1 & N@5 & N@10 & N@1 & N@5 & N@10& N@1 & N@5 & N@10 \\ 
% \cline{2-3}

\toprule
% \multirow{3}*{Traditional}&
% \rowcolor{mygray}
% \multicolumn{2}{l}{\textbf{Traditional recommender models:}}&&&&&&&&
% \\
Pop & 
0.26&
0.60&
0.63&
0.39&
0.61&
0.67&
0.78&
0.86&
0.88

 \\
BPRMF & 
0.44&
0.71&
0.74&
0.57&
0.75&
0.78&
0.71&
0.80&
0.84
 \\
LightGCN&
0.33&
0.62&
0.65&
0.47&
0.70&
0.73&
0.79
&0.82
&0.88\\

SASRec & 
\underline{0.68}&
\underline{0.84}&
\underline{0.85}&
\underline{0.69}&
\underline{0.83}&
\underline{0.85}&
\underline{0.81}&
\underline{0.88}&
\underline{0.89}
 \\
 \toprule
%  \rowcolor{mygray}
% \textbf{Zero-shot methods:}&&&&&&&&&\\
BM25 & 
0.08&
0.2&
0.43&
0.27&
0.45&
0.57&
-&
-&
-
 \\
 UniSRec & 
 0.12&
 0.35&
 0.47&
 0.25&
 0.45&
 0.56&
 -&
 -&
 -

 \\
VQ-Rec & 
0.10&
0.33&
0.47&
0.14&
0.33&
0.48&
-&
-&
-
 \\
 \toprule
% \rowcolor{mygray}
% \textbf{Prompt-based methods:}&&&&&&&&&\\
Manual prompts (task-wise)&
0.03\tiny{$\pm$0.02} &
0.35\tiny{$\pm$0.01}&
0.41\tiny{$\pm$0.01}&
0.04\tiny{$\pm$0.01}&
0.30\tiny{$\pm$0.02}&
0.35\tiny{$\pm$0.02}&
0.05\tiny{$\pm$0.03}&
0.39\tiny{$\pm$0.01}&
0.43\tiny{$\pm$0.01}\\

Enumeration (task-wise)&
0.04\tiny{$\pm$0.00}&
0.36\tiny{$\pm$0.00}&
0.42\tiny{$\pm$0.01}&
0.04\tiny{$\pm$0.00}&
0.34\tiny{$\pm$0.01}&
0.37\tiny{$\pm$0.01}&
0.05\tiny{$\pm$0.00}&
0.40\tiny{$\pm$0.01}&
0.44\tiny{$\pm$0.02}\\
GRIPS (instance-wise)&
0.43\tiny{$\pm$0.01}&
0.70\tiny{$\pm$0.03}&
0.73\tiny{$\pm$0.05}&
0.62\tiny{$\pm$0.04}& 
0.67\tiny{$\pm$0.02}&
0.72\tiny{$\pm$0.02}&
0.19\tiny{$\pm$0.03}&
0.64\tiny{$\pm$0.01}&
0.66\tiny{$\pm$0.04}
\\
\toprule
RPP (instance-wise)& 
 {\color{red}0.82\tiny{$\pm$0.03} }$ \uparrow$&
 {\color{red}0.85\tiny{$\pm$0.03}}$ \uparrow$&
 {\color{red}0.87\tiny{$\pm$0.02}}$ \uparrow$&
 {\color{red}0.79\tiny{$\pm$0.03}}$ \uparrow$&
 {0.82\tiny{$\pm$0.03}}$ \uparrow$&
 {\color{red}0.85\tiny{$\pm$0.02}}$ \uparrow$&
 {\color{red}0.87\tiny{$\pm$ 0.01}}$ \uparrow$&
 {\color{red}0.89\tiny{$\pm$0.01}}$ \uparrow$&
 {\color{red}0.91\tiny{$\pm$0.01}}$ \uparrow$
 \\
\rowcolor{mygray}Improvement& 
0.78&0.49&0.45&0.75&0.48&0.48&0.82&0.49&0.47\\
\rowcolor{mygray}Relative improvement (\%)& 
1950&136&107&1875&141&130&1640&123&107\\
 % \toprule
 RPP+ (instance-wise)& 
 {\color{red}0.82\tiny{$\pm$0.01} }$ \uparrow$&
 {\color{red}0.85\tiny{$\pm$0.00}}$ \uparrow$&
 {\color{red}0.87\tiny{$\pm$0.01}}$ \uparrow$&
 {\color{red}0.83\tiny{$\pm$0.02}}$ \uparrow$&
 {\color{red}0.85\tiny{$\pm$0.01}}$ \uparrow$&
 {\color{red}0.86\tiny{$\pm$0.01}}$ \uparrow$&
 {\color{red}0.93\tiny{$\pm$ 0.02}}$ \uparrow$&
 {\color{red}0.94\tiny{$\pm$0.01}}$ \uparrow$&
 {\color{red}0.95\tiny{$\pm$0.02}}$ \uparrow$\\
 \rowcolor{mygray}Improvement& 0.78&0.49&0.45&0.79&0.51&0.49&0.88&0.54&0.48
 \\
 \rowcolor{mygray}Relative improvement (\%)& 
1950&136&107&1975&150&132&1760&135&109\\
% \cmidrule(lr){3-11}

\bottomrule
% \vspace{-4mm}
\end{tabular}}
\end{table*}

\begin{table*}[htb]
\renewcommand\arraystretch{1.2}
 % \vspace{-4mm}
    \caption{{Additional metrics comparison of our framework on LLaMa2-7B-chat with baselines across diverse datasets, where ``M@'' means MRR metrics and ``H@'' means HitRatio metrics. The results of the best-performing baseline are denoted with an \underline{underline}. The results of our method which outperform other prompt-based methods are indicated with $\uparrow$}}
 % \vspace{-3mm}
\label{table:main_add_metrics}
\centering
\small
\setlength{\tabcolsep}{1.2mm}
{\begin{tabular}{llcccccccc}
\toprule
\multirow{2}*{methods} & \multicolumn{3}{c}{ML-1M} & \multicolumn{3}{c}{Games}&\multicolumn{3}{c}{Lastfm}  \\
\cmidrule(lr){2-4}\cmidrule(lr){5-7}\cmidrule(lr){8-10}
 &M@5 & M@10 & H@5 &M@5 & M@10 & H@5&M@5 & M@10 & H@5 \\ 
% \cline{2-3}

\toprule

SASRec & 
  
\underline{0.67}&
\underline{0.78}&
\underline{0.80}&
\underline{0.67}&
\underline{0.76}&
\underline{0.87}&
\underline{0.71}&
\underline{0.75}&
\underline{0.81}

 \\
 \toprule
Manual prompts (task-wise)&
0.24\tiny{$\pm$0.00} &
0.25\tiny{$\pm$0.01}&
0.69\tiny{$\pm$0.02}&
0.29\tiny{$\pm$0.01}&
0.39\tiny{$\pm$0.02}&
0.70\tiny{$\pm$0.02}&
0.27\tiny{$\pm$0.02}&
0.29\tiny{$\pm$0.02}&
0.76\tiny{$\pm$0.01}\\

Enumeration (task-wise)&
0.28\tiny{$\pm$0.02}&
0.29\tiny{$\pm$0.03}&
0.70\tiny{$\pm$0.01}&

0.38\tiny{$\pm$0.01}&
0.45\tiny{$\pm$0.01}&
0.72\tiny{$\pm$0.00}&

0.31\tiny{$\pm$0.01}&
0.32\tiny{$\pm$0.00}&
0.84\tiny{$\pm$0.01}\\

GRIPS (instance-wise)&
0.52\tiny{$\pm$0.03}&
0.54\tiny{$\pm$0.02}&
0.74\tiny{$\pm$0.02}&
0.51\tiny{$\pm$0.03}& 
0.52\tiny{$\pm$0.04}&
0.73\tiny{$\pm$0.02}&
0.50\tiny{$\pm$0.01}&
0.51\tiny{$\pm$0.03}&
0.76\tiny{$\pm$0.02}
\\
\toprule
RPP (instance-wise)& 
 {0.80\tiny{$\pm$0.01} }$ \uparrow$&
 {0.82\tiny{$\pm$0.03}}$ \uparrow$&
 {0.86\tiny{$\pm$0.03}}$ \uparrow$&
 {0.80\tiny{$\pm$0.02}}$ \uparrow$&
 {0.83\tiny{$\pm$0.02}}$ \uparrow$&
 {0.84\tiny{$\pm$0.03}}$ \uparrow$&
 {0.85\tiny{$\pm$ 0.01}}$ \uparrow$&
 {0.87\tiny{$\pm$0.02}}$ \uparrow$&
 {0.87\tiny{$\pm$0.02}}$ \uparrow$

 \\
 
 RPP+ (instance-wise)& 
{0.82\tiny{$\pm$0.02} }$ \uparrow$&
 {0.83\tiny{$\pm$0.01}}$ \uparrow$&
 {0.87\tiny{$\pm$0.00}}$ \uparrow$&
 {0.85\tiny{$\pm$0.01}}$ \uparrow$&
 {0.86\tiny{$\pm$0.00}}$ \uparrow$&
 {0.87\tiny{$\pm$0.01}}$ \uparrow$&
 {0.88\tiny{$\pm$ 0.01}}$ \uparrow$&
 {0.89\tiny{$\pm$0.00}}$ \uparrow$&
 {0.89\tiny{$\pm$0.01}}$ \uparrow$ \\

\bottomrule
% \vspace{-4mm}
\end{tabular}}
\end{table*}

\begin{table*}[htb]
 % \vspace{-1mm}
\caption{Performance of RPP and RPP+ on diverse frozen LLMs. The results of our framework which outperform other prompt-based baselines are indicated with $\uparrow$, and the best-performing methods are highlighted in {\color{red}{red}}. L, C, and A denote LLMs of LLaMa2-7B-chat, ChatGPT, and Alpaca, respectively. }
% \vspace{-3mm}
\label{table:genralization}
\centering
% \small
\renewcommand\arraystretch{1.2}
\setlength{\tabcolsep}{1.4mm}
{\begin{tabular}{lccccccccc}
\toprule
\multirow{2}*{methods} & \multicolumn{3}{c}{ML-1M} & \multicolumn{3}{c}{Games}&\multicolumn{3}{c}{Lastfm}  \\
\cmidrule(lr){2-4}\cmidrule(lr){5-7}\cmidrule(lr){8-10}
 &N@1 & N@5 & N@10 & N@1 & N@5 & N@10& N@1 & N@5 & N@10 \\ 

 \toprule
%  \rowcolor{mygray}
% \textbf{LLaMa2-7B-chat:}&&&&&&&&&\\

Manual prompts (L)&
0.03\tiny{$\pm$0.02} &
0.35\tiny{$\pm$0.01}&
0.41\tiny{$\pm$0.01}&
0.04\tiny{$\pm$0.01}&
0.30\tiny{$\pm$0.02}&
0.35\tiny{$\pm$0.02}&
0.05\tiny{$\pm$0.03}&
0.39\tiny{$\pm$0.01}&
0.43\tiny{$\pm$0.01}\\

Enumeration (L)&
0.04\tiny{$\pm$0.00}&
0.36\tiny{$\pm$0.00}&
0.42\tiny{$\pm$0.01}&
0.04\tiny{$\pm$0.00}&
0.34\tiny{$\pm$0.01}&
0.37\tiny{$\pm$0.01}&
0.05\tiny{$\pm$0.00}&
0.40\tiny{$\pm$0.01}&
0.44\tiny{$\pm$0.02}\\
GRIPS (L)&
0.43\tiny{$\pm$0.01}&
0.70\tiny{$\pm$0.03}&
0.73\tiny{$\pm$0.05}&
0.62\tiny{$\pm$0.04}& 
0.67\tiny{$\pm$0.02}&
0.72\tiny{$\pm$0.02}&
0.19\tiny{$\pm$0.03}&
0.64\tiny{$\pm$0.01}&
0.66\tiny{$\pm$0.04}
\\
RPP (L)& 
 {0.82\tiny{$\pm$0.03} }$ \uparrow$&
 {0.85\tiny{$\pm$0.03}}$ \uparrow$&
 {0.87\tiny{$\pm$0.02}}$ \uparrow$&
 {0.79\tiny{$\pm$0.03}}$ \uparrow$&
 {0.82\tiny{$\pm$0.03}}$ \uparrow$&
 {0.85\tiny{$\pm$0.02}}$ \uparrow$&
 {0.87\tiny{$\pm$ 0.01}}$ \uparrow$&
 {0.89\tiny{$\pm$0.01}}$ \uparrow$&
 {0.91\tiny{$\pm$0.01}}$ \uparrow$
 \\
RPP+ (L) & 
 {\color{red}0.82\tiny{$\pm$0.01} }$ \uparrow$&
 {\color{red}0.85\tiny{$\pm$0.00}}$ \uparrow$&
 {\color{red}0.87\tiny{$\pm$0.01}}$ \uparrow$&
 {\color{red}0.83\tiny{$\pm$0.02}}$ \uparrow$&
 {\color{red}0.85\tiny{$\pm$0.01}}$ \uparrow$&
 {\color{red}0.86\tiny{$\pm$0.01}}$ \uparrow$&
 {\color{red}0.93\tiny{$\pm$ 0.02}}$ \uparrow$&
 {\color{red}0.94\tiny{$\pm$0.01}}$ \uparrow$&
 {\color{red}0.95\tiny{$\pm$0.02}}$ \uparrow$
 \\
% \cmidrule(lr){3-11}
 \toprule
%   \rowcolor{mygray}
% \textbf{ChatGPT:}&&&&&&&&&\\
Manual prompts (C)&
 0.25\tiny{$\pm$0.03}&
 0.53\tiny{$\pm$0.03}&
 0.57\tiny{$\pm$0.03}&
 0.24\tiny{$\pm$0.06}&
 0.46\tiny{$\pm$0.05}&
 0.52\tiny{$\pm$0.05}&
 0.13\tiny{$\pm$0.09}&
 0.41\tiny{$\pm$0.05}&
 0.50\tiny{$\pm$0.06}\\

Enumeration (C)&
0.29\tiny{$\pm$0.00}&
0.55\tiny{$\pm$0.00}&
0.62\tiny{$\pm$0.01}&
0.31\tiny{$\pm$0.00}&
0.50\tiny{$\pm$0.01}&
0.56\tiny{$\pm$0.02}&
0.24\tiny{$\pm$0.02}&
0.47\tiny{$\pm$0.03}&
0.52\tiny{$\pm$0.02}\\
GRIPS (C)&
{\color{red}{0.41\tiny{$\pm$0.03}}}&
0.55\tiny{$\pm$0.02}&
0.65\tiny{$\pm$0.01}&
0.25\tiny{$\pm$0.02}& 
0.55\tiny{$\pm$0.01}&
0.62\tiny{$\pm$0.02}&
0.41\tiny{$\pm$0.02}&
0.64\tiny{$\pm$0.02}&
0.69\tiny{$\pm$0.04}
\\
RPP (C)&
 {0.35\tiny{$\pm$0.04}}$ \uparrow$&
 {0.58\tiny{$\pm$0.03}}$ \uparrow$&
 {0.65\tiny{$\pm$0.02}}$ \uparrow$&
 {0.53\tiny{$\pm$0.02}}$ \uparrow$&
 {0.67\tiny{$\pm$0.01}}$ \uparrow$&
 {0.73\tiny{$\pm$0.01}}$ \uparrow$&
 {0.66\tiny{$\pm$0.02}}$ \uparrow$&
 {0.83\tiny{$\pm$0.02}}$ \uparrow$&
 {0.83\tiny{$\pm$0.01}}$ \uparrow$\\
RPP+ (C)&
 {0.38\tiny{$\pm$0.07}}$ \uparrow$&
 {\color{red}0.59\tiny{$\pm$0.02}}$ \uparrow$&
 {\color{red}0.66\tiny{$\pm$0.01}}$ \uparrow$&
 {\color{red}0.59\tiny{$\pm$0.02}}$ \uparrow$&
 {\color{red}0.69\tiny{$\pm$0.02}}$ \uparrow$&
 {\color{red}0.75\tiny{$\pm$0.01}}$ \uparrow$&
 {\color{red}0.67\tiny{$\pm$0.02}}$ \uparrow$&
 {\color{red}0.84\tiny{$\pm$0.01}}$ \uparrow$&
 {\color{red}0.85\tiny{$\pm$0.01}}$ \uparrow$\\
 \toprule
% \cmidrule(lr){3-11}
% \rowcolor{mygray}
% \textbf{Fine-tuned Alpaca:}&&&&&&&&&\\
Manual prompts (A)&
0.04\tiny{$\pm$0.04}&
0.24\tiny{$\pm$0.16}&
0.34\tiny{$\pm$0.04}&
0.09\tiny{$\pm$0.02}&
0.34\tiny{$\pm$0.04}&
0.43\tiny{$\pm$0.02}&
0.06\tiny{$\pm$0.06}&
0.29\tiny{$\pm$0.16}&
0.40\tiny{$\pm$0.05}
\\
Enumeration (A)&
0.06\tiny{$\pm$0.01}&
0.30\tiny{$\pm$0.02}&
0.37\tiny{$\pm$0.02}&
0.14\tiny{$\pm$0.01}&
0.39\tiny{$\pm$0.01}&
0.43\tiny{$\pm$0.02}&
0.12\tiny{$\pm$0.00}&
0.38\tiny{$\pm$0.01}&
0.44\tiny{$\pm$0.01}\\
GRIPS (A)&
0.06\tiny{$\pm$0.01}&
0.63\tiny{$\pm$0.02}&
0.65\tiny{$\pm$0.00}&
0.02\tiny{$\pm$0.01}&
0.60\tiny{$\pm$0.02}&
0.63\tiny{$\pm$0.02}&
0.19\tiny{$\pm$0.02}&
0.68\tiny{$\pm$0.01}&
0.70\tiny{$\pm$0.00}
\\
RPP (A)&
 { 0.82\tiny{$\pm$0.1}}$ \uparrow$&
  {0.92\tiny{$\pm$0.01}}$ \uparrow$&
  {0.93\tiny{$\pm$0.03}}$ \uparrow$&
  {0.96\tiny{$\pm$0.00}}$ \uparrow$&
  {0.96\tiny{$\pm$0.00}}$ \uparrow$&
  {0.97\tiny{$\pm$0.00}}$ \uparrow$&
  {0.91\tiny{$\pm$0.03}}$ \uparrow$&
  {0.91\tiny{$\pm$0.03}}$ \uparrow$&
  {\color{red}0.93\tiny{$\pm$0.01}}$ \uparrow$\\
RPP+ (A) & 
 {\color{red}0.84\tiny{$\pm$0.12} }$ \uparrow$&
 {\color{red}0.94\tiny{$\pm$0.05}}$ \uparrow$&
 {\color{red}0.94\tiny{$\pm$0.05}}$ \uparrow$&
 {\color{red}0.96\tiny{$\pm$0.00}}$ \uparrow$&
 {\color{red}0.96\tiny{$\pm$0.01}}$ \uparrow$&
 {\color{red}0.97\tiny{$\pm$0.00}}$ \uparrow$&
 {\color{red}0.92\tiny{$\pm$ 0.00}}$ \uparrow$&
 {\color{red}0.92\tiny{$\pm$0.00}}$ \uparrow$&
 {0.92\tiny{$\pm$0.00}}$ \uparrow$
 \\
\bottomrule
% \vspace{-4mm}
\end{tabular}}
\end{table*}

\begin{table*}[htb]
\renewcommand\arraystretch{1.2}
% \vspace{-1mm}
\caption{Ablation study on the respective impacts of four patterns on RPP.}
% \vspace{-3mm}
\label{table: ablation}
\centering
% \small

% \renewcommand\arraystretch{1.05}

\setlength{\tabcolsep}{6pt}
{\begin{tabular}{lccccccccc}
\toprule
\multirow{2}*{methods} & \multicolumn{3}{c}{ML-1M} & \multicolumn{3}{c}{Games}&\multicolumn{3}{c}{Lastfm}  \\
\cmidrule(lr){2-4}\cmidrule(lr){5-7}\cmidrule(lr){8-10}
 &N@1 & N@5 & N@10 & N@1 & N@5 & N@10& N@1 & N@5 & N@10\\ 
% \cline{2-3}

\toprule
Manual prompts &
0.03\tiny{$\pm$0.02} &
0.35\tiny{$\pm$0.01}&
0.41\tiny{$\pm$0.01}&
0.04\tiny{$\pm$0.01}&
0.30\tiny{$\pm$0.02}&
0.35\tiny{$\pm$0.02}&
0.05\tiny{$\pm$0.03}&
0.39\tiny{$\pm$0.01}&
0.43\tiny{$\pm$0.01}\\

Enumeration&
0.04\tiny{$\pm$0.00}&
0.36\tiny{$\pm$0.00}&
0.42\tiny{$\pm$0.01}&
0.04\tiny{$\pm$0.00}&
0.34\tiny{$\pm$0.01}&
0.37\tiny{$\pm$0.01}&
0.05\tiny{$\pm$0.00}&
0.40\tiny{$\pm$0.01}&
0.44\tiny{$\pm$0.02}\\
GRIPS &
0.43\tiny{$\pm$0.01}&
0.70\tiny{$\pm$0.03}&
0.73\tiny{$\pm$0.05}&
0.62\tiny{$\pm$0.04}& 
0.67\tiny{$\pm$0.02}&
0.72\tiny{$\pm$0.02}&
0.19\tiny{$\pm$0.03}&
0.64\tiny{$\pm$0.01}&
0.66\tiny{$\pm$0.04}
\\
 \toprule

Role-playing & 
0.76\tiny{$\pm$0.01}&
0.80\tiny{$\pm$0.01}&
0.84\tiny{$\pm$0.01}&
0.66\tiny{$\pm$0.05}&
0.72\tiny{$\pm$0.05}&
0.76\tiny{$\pm$0.04}&
0.84\tiny{$\pm$0.02}&
0.87\tiny{$\pm$0.02}&
0.89\tiny{$\pm$0.02}
 \\
 History records& 
0.65\tiny{$\pm$0.10}&
0.72\tiny{$\pm$0.07}&
0.77\tiny{$\pm$0.06}&
0.67\tiny{$\pm$0.00}&
0.71\tiny{$\pm$0.01}&
0.76\tiny{$\pm$0.00}&
0.81\tiny{$\pm$0.02}&
0.85\tiny{$\pm$0.01}&
0.87\tiny{$\pm$0.01}
 \\
Reasoning guidance & 
0.78\tiny{$\pm$0.01}&
0.81\tiny{$\pm$0.01}&
0.84\tiny{$\pm$0.00}&
0.74\tiny{$\pm$0.04}&
0.78\tiny{$\pm$0.04}&
0.80\tiny{$\pm$0.04}&
0.85\tiny{$\pm$0.02}&
0.88\tiny{$\pm$0.02}&
0.89\tiny{$\pm$0.02}
 \\

Output format & 
 0.74\tiny{$\pm$0.00}&
 0.77\tiny{$\pm$0.04}&
 0.82\tiny{$\pm$0.03}&
 0.67\tiny{$\pm$0.02}&
 0.71\tiny{$\pm$0.02}&
 0.76\tiny{$\pm$0.02}&
 0.83\tiny{$\pm$0.03}&
 0.86\tiny{$\pm$0.04}&
 0.89\tiny{$\pm$0.03}
 \\
\toprule
RPP&
0.82\tiny{$\pm$0.03}&
0.85\tiny{$\pm$0.03}&
0.87\tiny{$\pm$0.02}&
0.79\tiny{$\pm$0.03}&
0.82\tiny{$\pm$0.03}&
0.85\tiny{$\pm$0.02}&
0.87\tiny{$\pm$ 0.01}&
0.89\tiny{$\pm$0.01}&
0.91\tiny{$\pm$0.01}\\

% Persona+ & 
% 0.75\tiny{$\pm$0.01}&
% 0.82\tiny{$\pm$0.03}&
% 0.84\tiny{$\pm$0.01}&
% 0.74\tiny{$\pm$0.04}&
% 0.79\tiny{$\pm$0.02}&
% 0.82\tiny{$\pm$0.04}&
% 0.86\tiny{$\pm$0.02}&
% 0.89\tiny{$\pm$0.02}&
% 0.90\tiny{$\pm$0.01}
%  \\
%  Records+& 
% 0.67\tiny{$\pm$0.08}&
% 0.73\tiny{$\pm$0.06}&
% 0.78\tiny{$\pm$0.03}&
% 0.66\tiny{$\pm$0.00}&
% 0.74\tiny{$\pm$0.01}&
% 0.81\tiny{$\pm$0.01}&
% 0.79\tiny{$\pm$0.04}&
% 0.86\tiny{$\pm$0.01}&
% 0.89\tiny{$\pm$0.03}
%  \\
% Recipe+ & 
% 0.78\tiny{$\pm$0.01}&
% 0.83\tiny{$\pm$0.03}&
% 0.85\tiny{$\pm$0.04}&
% 0.74\tiny{$\pm$0.05}&
% 0.78\tiny{$\pm$0.03}&
% 0.81\tiny{$\pm$0.03}&
% 0.83\tiny{$\pm$0.02}&
% 0.84\tiny{$\pm$0.04}&
% 0.88\tiny{$\pm$0.02}
%  \\

% Template+ & 
%  0.72\tiny{$\pm$0.01}&
%  0.77\tiny{$\pm$0.02}&
%  0.80\tiny{$\pm$0.03}&
%  0.76\tiny{$\pm$0.04}&
%  0.79\tiny{$\pm$0.03}&
%  0.80\tiny{$\pm$0.02}&
%  0.84\tiny{$\pm$0.05}&
%  0.87\tiny{$\pm$0.04}&
%  0.91\tiny{$\pm$0.03}
%  \\
%   % \toprule
% RPP+&
%  {0.82\tiny{$\pm$0.01} }&
%  {0.85\tiny{$\pm$0.00}}&
%  {0.87\tiny{$\pm$0.01}}&
%  {0.83\tiny{$\pm$0.02}}&
%  {0.85\tiny{$\pm$0.01}}&
%  {0.86\tiny{$\pm$0.01}}&
%  {0.93\tiny{$\pm$ 0.02}}&
%  {0.94\tiny{$\pm$0.01}}&
%  {0.95\tiny{$\pm$0.02}}\\
 
\bottomrule
\end{tabular}}
\end{table*}

\begin{table*}[htb]
\renewcommand\arraystretch{1.2}
% \vspace{-4mm}
\caption{Ablation study on the respective impacts of four patterns on RPP+.}
% \vspace{-3mm}
\label{table: ablation_rpp}
\centering
% \small

% \renewcommand\arraystretch{1.05}

\setlength{\tabcolsep}{6pt}
{\begin{tabular}{lccccccccc}
\toprule
\multirow{2}*{methods} & \multicolumn{3}{c}{ML-1M} & \multicolumn{3}{c}{Games}&\multicolumn{3}{c}{Lastfm}  \\
\cmidrule(lr){2-4}\cmidrule(lr){5-7}\cmidrule(lr){8-10}
 &N@1 & N@5 & N@10 & N@1 & N@5 & N@10& N@1 & N@5 & N@10\\ 
% \cline{2-3}

\toprule
Manual prompts &
0.03\tiny{$\pm$0.02} &
0.35\tiny{$\pm$0.01}&
0.41\tiny{$\pm$0.01}&
0.04\tiny{$\pm$0.01}&
0.30\tiny{$\pm$0.02}&
0.35\tiny{$\pm$0.02}&
0.05\tiny{$\pm$0.03}&
0.39\tiny{$\pm$0.01}&
0.43\tiny{$\pm$0.01}\\

Enumeration&
0.04\tiny{$\pm$0.00}&
0.36\tiny{$\pm$0.00}&
0.42\tiny{$\pm$0.01}&
0.04\tiny{$\pm$0.00}&
0.34\tiny{$\pm$0.01}&
0.37\tiny{$\pm$0.01}&
0.05\tiny{$\pm$0.00}&
0.40\tiny{$\pm$0.01}&
0.44\tiny{$\pm$0.02}\\
GRIPS &
0.43\tiny{$\pm$0.01}&
0.70\tiny{$\pm$0.03}&
0.73\tiny{$\pm$0.05}&
0.62\tiny{$\pm$0.04}& 
0.67\tiny{$\pm$0.02}&
0.72\tiny{$\pm$0.02}&
0.19\tiny{$\pm$0.03}&
0.64\tiny{$\pm$0.01}&
0.66\tiny{$\pm$0.04}
\\
 \toprule
Role-playing+ & 
0.75\tiny{$\pm$0.01}&
0.82\tiny{$\pm$0.03}&
0.84\tiny{$\pm$0.01}&
0.74\tiny{$\pm$0.04}&
0.79\tiny{$\pm$0.02}&
0.82\tiny{$\pm$0.04}&
0.86\tiny{$\pm$0.02}&
0.89\tiny{$\pm$0.02}&
0.90\tiny{$\pm$0.01}
 \\
History records+& 
0.67\tiny{$\pm$0.08}&
0.73\tiny{$\pm$0.06}&
0.78\tiny{$\pm$0.03}&
0.66\tiny{$\pm$0.00}&
0.74\tiny{$\pm$0.01}&
0.81\tiny{$\pm$0.01}&
0.79\tiny{$\pm$0.04}&
0.86\tiny{$\pm$0.01}&
0.89\tiny{$\pm$0.03}
 \\
Reasoning guidance+ & 
0.78\tiny{$\pm$0.01}&
0.83\tiny{$\pm$0.03}&
0.85\tiny{$\pm$0.04}&
0.74\tiny{$\pm$0.05}&
0.78\tiny{$\pm$0.03}&
0.81\tiny{$\pm$0.03}&
0.83\tiny{$\pm$0.02}&
0.84\tiny{$\pm$0.04}&
0.88\tiny{$\pm$0.02}
 \\

Output format+ & 
 0.72\tiny{$\pm$0.01}&
 0.77\tiny{$\pm$0.02}&
 0.80\tiny{$\pm$0.03}&
 0.76\tiny{$\pm$0.04}&
 0.79\tiny{$\pm$0.03}&
 0.80\tiny{$\pm$0.02}&
 0.84\tiny{$\pm$0.05}&
 0.87\tiny{$\pm$0.04}&
 0.91\tiny{$\pm$0.03}
 \\
  \toprule
RPP+&
 {0.82\tiny{$\pm$0.01} }&
 {0.85\tiny{$\pm$0.00}}&
 {0.87\tiny{$\pm$0.01}}&
 {0.83\tiny{$\pm$0.02}}&
 {0.85\tiny{$\pm$0.01}}&
 {0.86\tiny{$\pm$0.01}}&
 {0.93\tiny{$\pm$ 0.02}}&
 {0.94\tiny{$\pm$0.01}}&
 {0.95\tiny{$\pm$0.02}}\\
 
\bottomrule
% \vspace{-8mm}
\end{tabular}}
\end{table*}

\begin{table*}[htb]
\renewcommand\arraystretch{1.2}
 % \vspace{-4mm}
    \caption{{Additional experiments for the pattern of ``user profile''. }}
 % \vspace{-3mm}
\label{table:main_profile}
\centering
\small
\setlength{\tabcolsep}{1.2mm}
{\begin{tabular}{lccc}
\toprule
\multirow{2}*{methods} & \multicolumn{3}{c}{ML-1M}  \\
\cmidrule(lr){2-4}
 &N@1 & N@5 & N@10 \\ 
% \cline{2-3}

\toprule

Manual prompts (task-wise)&
0.05\tiny{$\pm$0.01}&
0.41\tiny{$\pm$0.02}&
0.46\tiny{$\pm$0.01}
\\

Enumeration (task-wise)&
0.16\tiny{$\pm$0.00} &
0.48\tiny{$\pm$0.02}&
0.51\tiny{$\pm$0.02}
\\

GRIPS (instance-wise)&
0.63\tiny{$\pm$0.01}&
0.72\tiny{$\pm$0.02}&
0.77\tiny{$\pm$0.00}
\\
\toprule
user profile (instance-wise)& 
 {0.78\tiny{$\pm$0.01} }$ \uparrow$&
 {0.79\tiny{$\pm$0.03}}$ \uparrow$&
 {0.82\tiny{$\pm$0.03}}$ \uparrow$
 \\
RPP (instance-wise)& 
 {0.83\tiny{$\pm$0.02} }$ \uparrow$&
 {0.83\tiny{$\pm$0.01}}$ \uparrow$&
 {0.88\tiny{$\pm$0.01}}$ \uparrow$
 \\
 
 RPP+ (instance-wise)& 
{0.83\tiny{$\pm$0.00} }$ \uparrow$&
 {0.84\tiny{$\pm$0.01}}$ \uparrow$&
 {0.89\tiny{$\pm$0.00}}$ \uparrow$
\\

\bottomrule
% \vspace{-4mm}
\end{tabular}}
\end{table*}

\begin{table*}[htb]
\renewcommand\arraystretch{1.2}
\caption{Ablation study on the impact of GRU.}
% \vspace{-3mm}
\label{table: ablation_gru}
\centering
\setlength{\tabcolsep}{6pt}
{\begin{tabular}{lccccccccc}
\toprule
\multirow{2}*{methods} & \multicolumn{3}{c}{ML-1M} & \multicolumn{3}{c}{Games}&\multicolumn{3}{c}{Lastfm}  \\
\cmidrule(lr){2-4}\cmidrule(lr){5-7}\cmidrule(lr){8-10}
 &N@1 & N@5 & N@10 & N@1 & N@5 & N@10& N@1 & N@5 & N@10\\ 
% \cline{2-3}

\toprule
Manual prompts &
0.03\tiny{$\pm$0.02} &
0.35\tiny{$\pm$0.01}&
0.41\tiny{$\pm$0.01}&
0.04\tiny{$\pm$0.01}&
0.30\tiny{$\pm$0.02}&
0.35\tiny{$\pm$0.02}&
0.05\tiny{$\pm$0.03}&
0.39\tiny{$\pm$0.01}&
0.43\tiny{$\pm$0.01}\\

Enumeration&
0.04\tiny{$\pm$0.00}&
0.36\tiny{$\pm$0.00}&
0.42\tiny{$\pm$0.01}&
0.04\tiny{$\pm$0.00}&
0.34\tiny{$\pm$0.01}&
0.37\tiny{$\pm$0.01}&
0.05\tiny{$\pm$0.00}&
0.40\tiny{$\pm$0.01}&
0.44\tiny{$\pm$0.02}\\
GRIPS &
0.43\tiny{$\pm$0.01}&
0.70\tiny{$\pm$0.03}&
0.73\tiny{$\pm$0.05}&
0.62\tiny{$\pm$0.04}& 
0.67\tiny{$\pm$0.02}&
0.72\tiny{$\pm$0.02}&
0.19\tiny{$\pm$0.03}&
0.64\tiny{$\pm$0.01}&
0.66\tiny{$\pm$0.04}
\\
 \toprule
RPP-pool&
 {0.80\tiny{$\pm$0.00} }&
 {0.84\tiny{$\pm$0.01}}&
 {0.86\tiny{$\pm$0.02}}&
 {0.76\tiny{$\pm$0.03}}&
 {0.79\tiny{$\pm$0.00}}&
 {0.81\tiny{$\pm$0.02}}&
 {0.90\tiny{$\pm$ 0.01}}&
 {0.92\tiny{$\pm$0.01}}&
 {0.93\tiny{$\pm$0.00}}\\
RPP-Bert&
 {0.82\tiny{$\pm$0.01} }&
 {0.85\tiny{$\pm$0.00}}&
 {0.87\tiny{$\pm$0.01}}&
 {0.83\tiny{$\pm$0.02}}&
 {0.85\tiny{$\pm$0.01}}&
 {0.86\tiny{$\pm$0.01}}&
 {0.93\tiny{$\pm$ 0.02}}&
 {0.94\tiny{$\pm$0.01}}&
 {0.95\tiny{$\pm$0.02}}\\
 
\bottomrule
% \vspace{-8mm}
\end{tabular}}
\end{table*}

\subsection{Overall Results (\textbf{RQ1})}
To answer RQ1, we evaluate the effectiveness of our framework on LLaMa2-7B-chat by comparing the performance of RPP/RPP+ with diverse baselines. We adopt the evaluating metrics of NDCG@1, NDCG@5, and NDCG@10 to access LLMs' capability in ranking items, the results are listed in Table~\ref{table:main_all}. Our observations are as follows: 1) Task-wise prompting methods (\ie manual prompts and enumeration) exhibit inferior recommendation performance compared to traditional recommender models, few-shot methods, and instance-wise prompting methods (\ie GRIPS  and RPP/RPP+). For example, the best-performing baseline in traditional recommender models, SASRec, surpasses the best-performing task-wise prompting method (enumeration) by $0.64$, $0.48$, and $0.43$ in NDCG@1, NDCG@5, and NDCG@10 metrics, respectively, on the ML-1M dataset. This validates the weakness of task-wise prompts for LLMs. 2) Instance-wise prompting implemented by RPP/RPP+ makes the frozen pre-trained LLaMa2-7B-chat surpass traditional recommender models across a variety of datasets. For instance, RPP demonstrates superior performance compared to the leading baseline in traditional recommender models (SASRec), achieving improvements of $0.14$, $0.01$, and $0.02$ in terms of the NDCG@1, NDCG@5, and NDCG@10 metrics on the ML-1M. 3) The framework of RPP/RPP+ outperforms other instance-wise promoting methods in the ranking tasks, such as GRIPS. This proves the high quality of prompts generated by RPP/RPP+. We conduct multiple tests and the minimal standard deviations demonstrate that the performance improvements resulting from RPP/RPP+ are statistically significant. {Additionally, we adopt the MRR@5, MRR@10, and HitRatio@5 metrics to evaluate the performance of RPP/RPP+ in Table \ref{table:main_add_metrics}, as these are widely used benchmarks. The results provide valuable insights that highlight the superiority of our method, which achieves state-of-the-art performance compared to the baselines.}
These overall results illustrate that due to the ability to personalize prompts from instance-wise, RPP/RPP+ enables LLMs to comprehend users' personalized intents better and provide more accurate recommendations.

\subsection{Generalization ability (\textbf{RQ2})}
To answer RQ2, we extend the evaluation of RPP/RPP+ on diverse frozen LLMs (\ie{LLaMa2-7B-chat, ChatGPT, and Alpaca}) and compare the performance of RPP/RPP+ with other prompt-based baselines. The results are listed in Table~\ref{table:genralization}. RPP/RPP+ can generalize well on diverse LLMs and prominently enhance the ranking performance of LLMs. 
For example, RPP outperforms enumeration in terms of the NDCG@10 metric on LLaMa-7B-chat, ChatGPT, and fine-tuned Alpaca by $0.48$, $0.24$, and $0.53$, respectively, on the Lastfm dataset.
The relatively less significant impact on ChatGPT's performance might be attributed to its stronger robustness to prompts compared with other LLMs. 
The consistent performance improvement across the three types of frozen LLMs demonstrates the effectiveness and generalization ability of RPP/RPP+. 
% We conduct multiple tests and the p-value<0.05 demonstrates that the performance improvements resulting from RPP/RPP+ are statistically significant. 

The results for Q1 and Q2 underscore the importance of instance-wise prompting and the effectiveness of RPP/RPP+ from a practical standpoint, confirming our viewpoint that personalizing instance-wise prompts can enhance LLMs' recommendation ability.
Furthermore, dynamically refining actions during the iterative process can lead to a slight performance improvement of RPP+ compared to RPP, possibly due to more flexible and higher-quality prompts after refinement. However, the improvement is not particularly significant {ranging from 0.00 to 0.06 and requires additional computational resources during the LLM refinement process,} the trade-off between efficiency and performance can be considered when choosing between the RPP and RPP+ in practice.

\begin{figure*}[t]
\centering  
\subfigure[The performance comparison of different patterns for RPP.]{
\label{Fig.ab1}
\includegraphics[width=6.5cm,height =5cm]{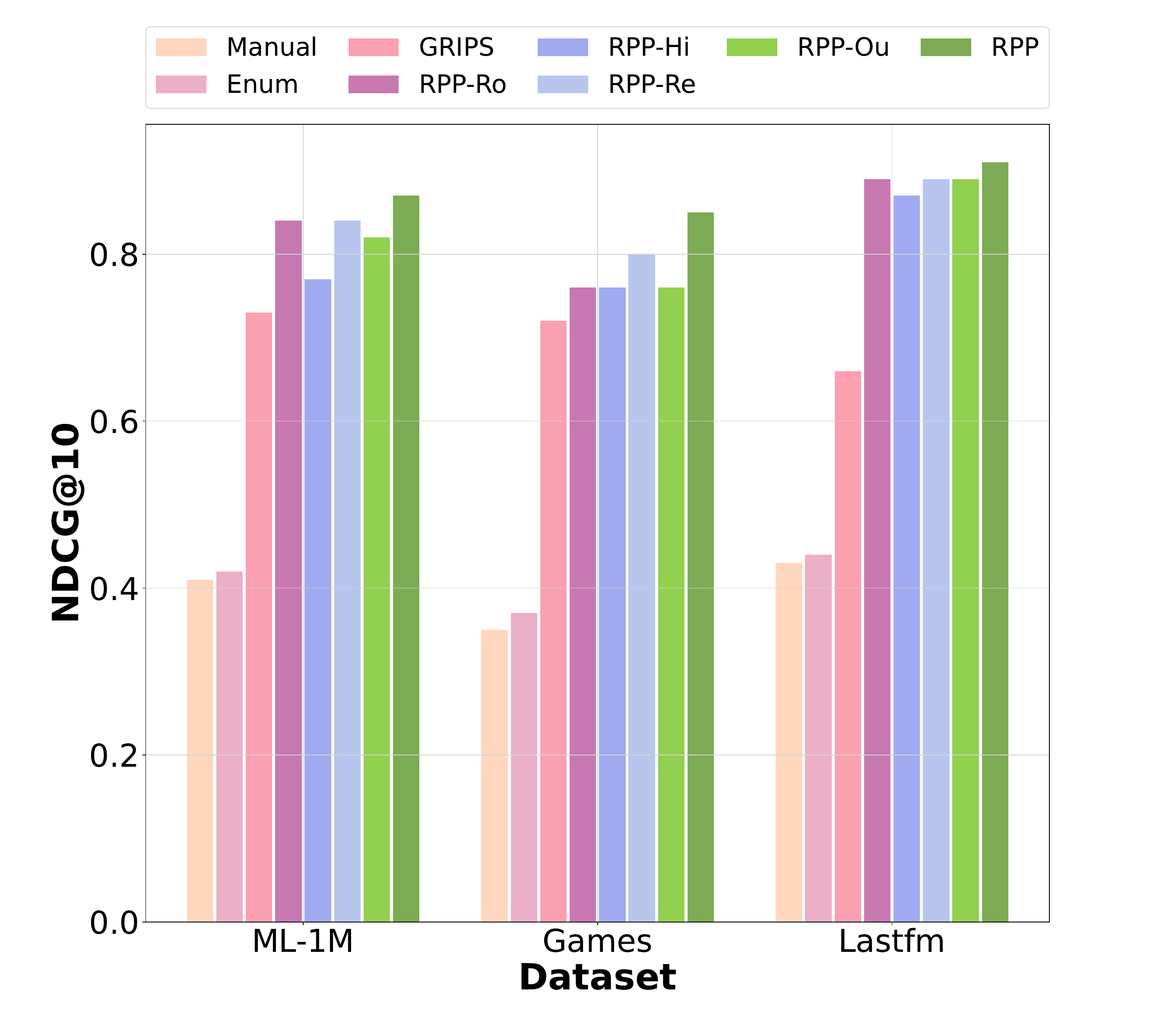}}\subfigure[The performance comparison of different patterns for RPP+.]{
\label{Fig.ab2}
\includegraphics[width=6.5cm,height =5cm]{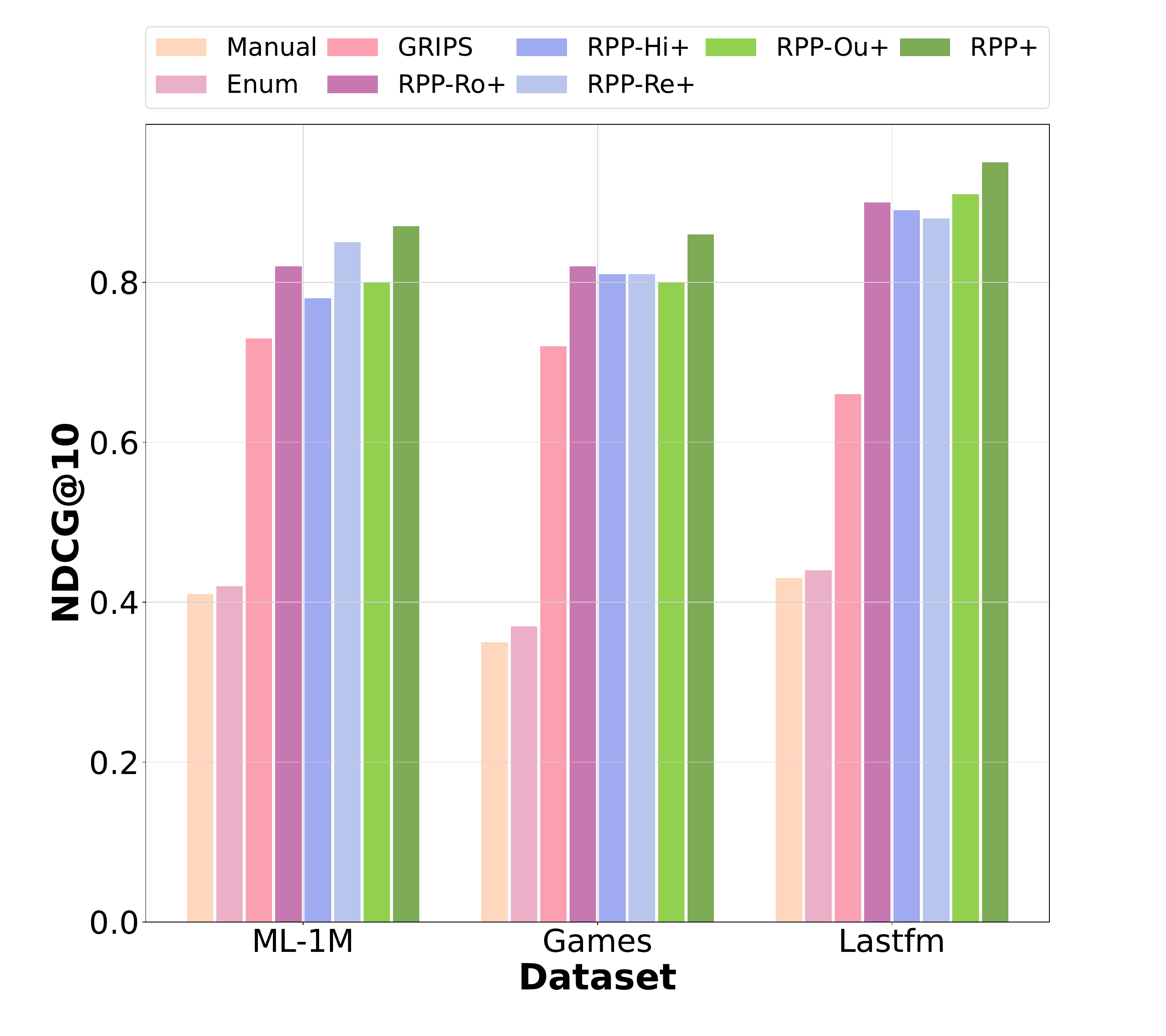}}
%\vspace{-6mm}
\caption{The performance comparison between prompt-based methods and RPP/RPP+ with different patterns. ``Manual'', ``Enum'', and ``GRIPS'' represent the baseline prompt-based methods. ``RPP-Ro'', ``RPP-Hi'', ``RPP-Re'', and ``RPP-Ou'' are the 4 variations of RPP/RPP+ on ``role-playing'', ``history records'', ``reasoning guidance'', and ``output format'' patterns, respectively.
}
\label{ab}
% \vspace{-2mm}
\end{figure*}

\begin{figure*}[t]
    \centering
    % \vspace{-5mm}
    \includegraphics[width=0.95\textwidth]{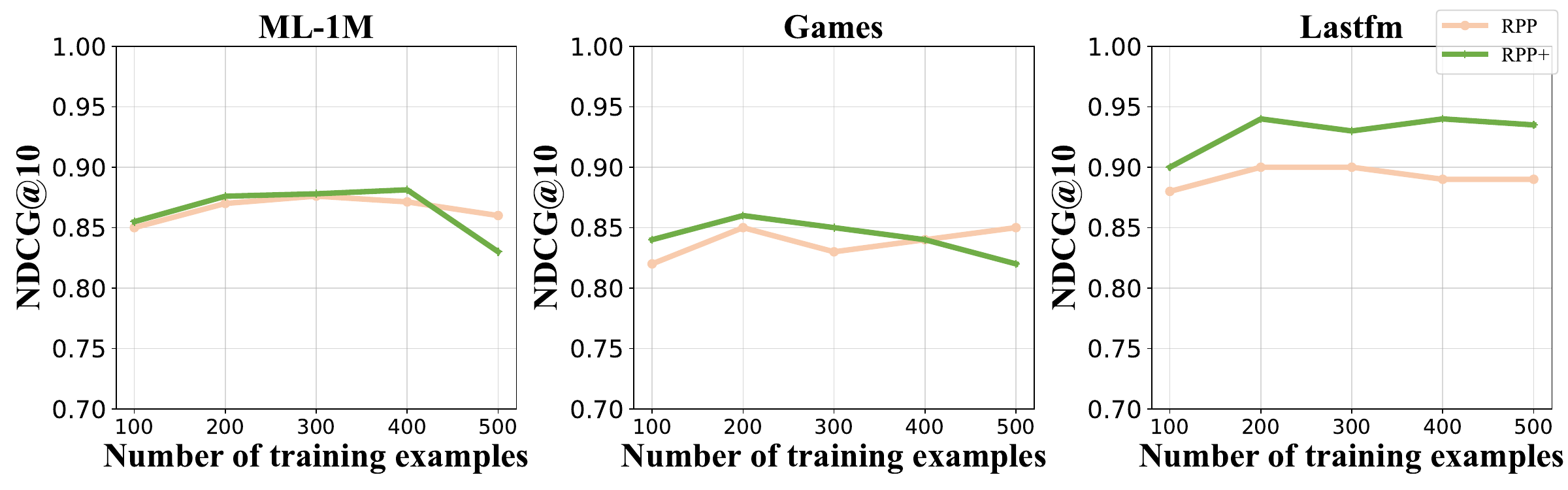}
     % \vspace{-8mm}
    \caption{Sensitivity to the number of training examples, which demonstrates the changes in RPP/RPP+'s performance on three datasets as the number of training examples increases from $100$ to $500$.}
    \label{Fig.sub.1}
    % \vspace{-2mm}
\end{figure*}

\begin{figure*}[t]
    \centering
    % \vspace{-5mm}
    \includegraphics[width=0.95\textwidth]{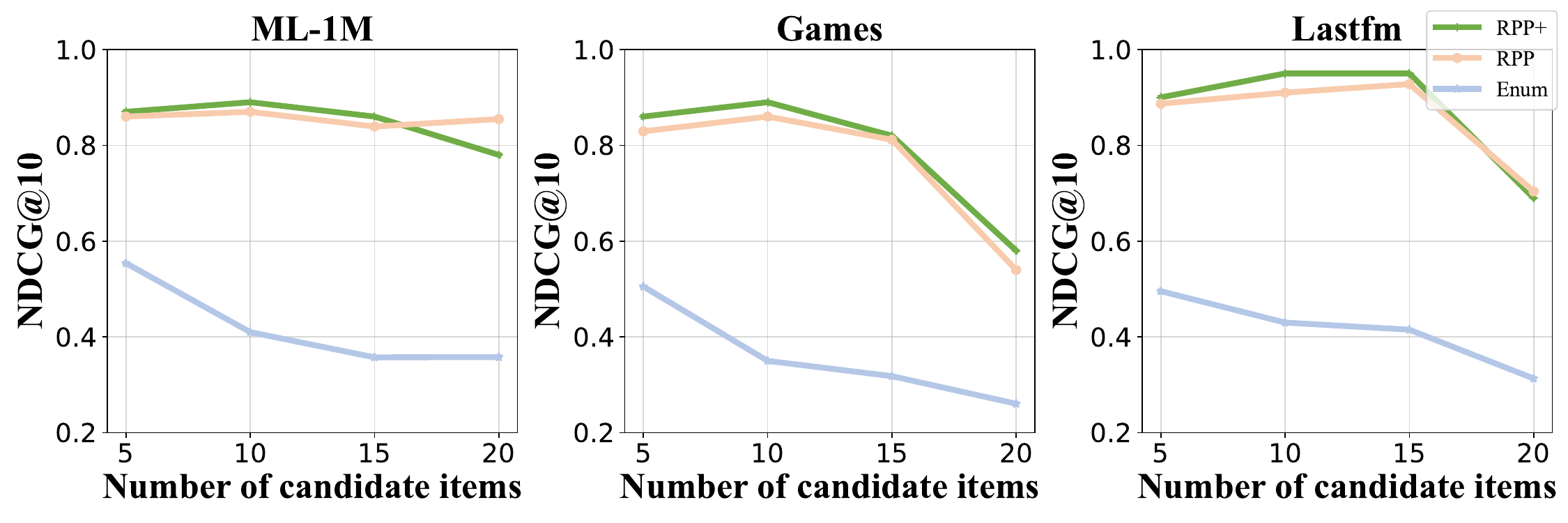}
     % \vspace{-8mm}
    \caption{Sensitivity to the number of candidate items, which indicates the drop in performance of LLM-based methods as the number of candidate items increases.}
    \label{Fig.sub.2}
    % \vspace{-2mm}
\end{figure*}

\begin{figure*}[t]
\centering  
\subfigure[{Ablation study for multi-agent reinforcement learning.}]{
\label{Fig.PPO}
\includegraphics[width=6.5cm,height =5cm]{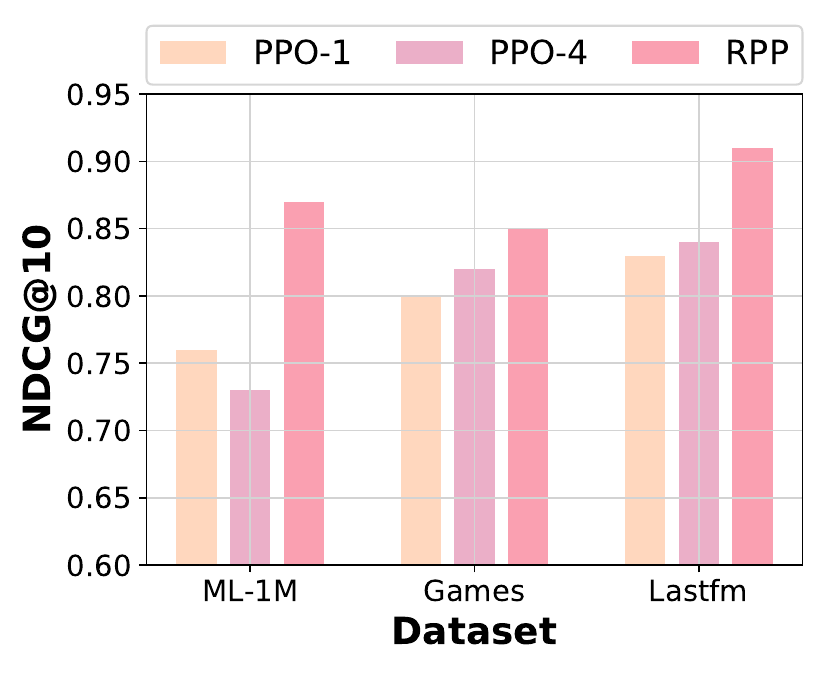}}
\subfigure[{Sensitivity to the number of sentence actions for each pattern.}]{
\label{Fig.num_action}
\includegraphics[width=6.5cm,height =5cm]{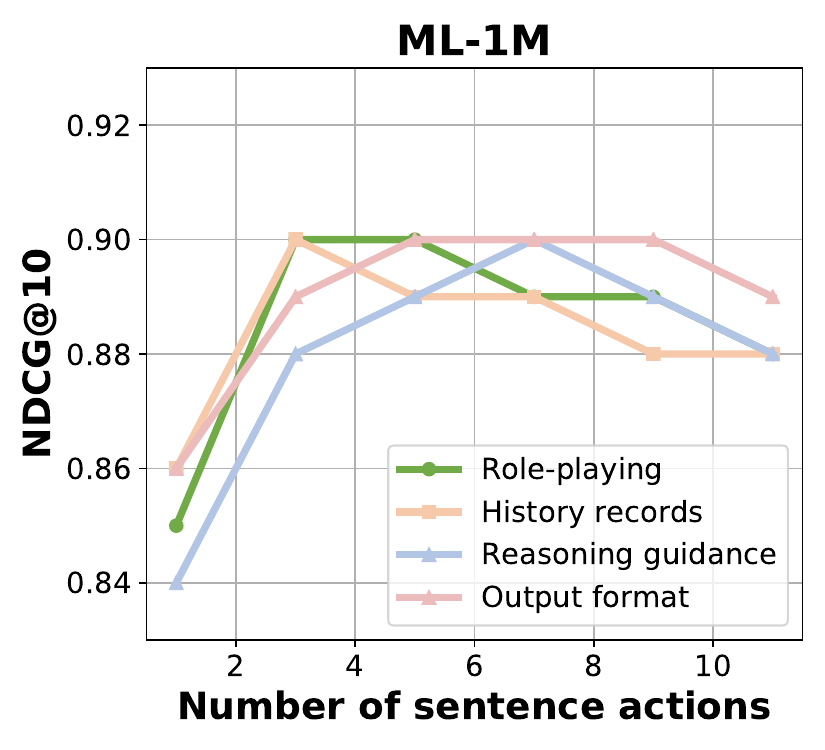}}
\vspace{-2mm}
\caption{{Figure (a) is the performance comparison between multi-agent reinforcement learning and a single PPO for personalizing prompts. Figure (b) demonstrates RPP/RPP+'s sensitivity to the number of actions under each pattern.}
}
\label{additional}
% \vspace{-2mm}
\end{figure*}

\subsection{Ablation Study (\textbf{RQ3})}
The ablation study explores the impact of personalizing each pattern (\ie{``role-playing'', ``history records'', ``reasoning guidance,'' and ``output format''}) on facilitating LLMs' recommendation ability. Specifically, we perform experiments on the frozen LLaMa2-7B-chat with 4 variations of RPP/RPP+, where only one pattern in the prompts is personalized by the corresponding agent. As shown in Table~\ref{table: ablation}, ~\ref{table: ablation_rpp}, and Figure~\ref{ab}, personalizing any of the patterns can enhance the performance of LLMs to rank items, and the full framework of RPP/RPP+ attains the highest performance. Personalizing ``reasoning guidance'' yields the greatest benefits compared with other patterns, providing interpretability for recommendations and demonstrating the effectiveness of the personalized reasoning process in accurately analyzing user preferences. Besides, ``role-playing'' and ``output format'' underscore LLMs' inherent sensitivity to prompts that modifications to the prompt expression can lead to LLMs' better output. The ``history records'' confirms the impact of users' short or long-term interaction history on recommendations.
This emphasizes that decomposing the prompt into four patterns and individually tailoring them can contribute to LLMs' recommendations.
{Furthermore, as some recommendation tasks focus on user modeling, we define a new pattern called "user profile" and conduct experiments to evaluate the effectiveness of RPP/RPP+ in personalizing this pattern using the ML-1M dataset. We utilize users' profile features, such as gender, age, and occupation, as potential actions for the agent, which can choose whether to include each feature in the prompts. As shown in Table \ref{table:main_profile}, personalizing the user profile leads to performance improvements compared with prompt-based baselines, highlighting the effectiveness of designating "user profile" as a distinct pattern and the advantages of personalizing prompts with RPP/RPP+.}

{To validate the design of ``GRU'' in Equation \ref{eq: S_t}, we conduct ablation experiments to replace ``GRU'' as ``mean pooling''. As presented in table \ref{table: ablation_gru}, the performance of ``RPP-pool'' is worse than that of ``RPP-Bert'', which validates the effectiveness of the ``GRU'' module since it can catch the sequential information of ranking results. Additionally, to evaluate the effectiveness of MARL which can personalize different patterns with different expertise from different agents, we utilize a single agent which is designed as PPO. The results are demonstrated in Figure \ref{Fig.PPO}, where the PPO-1 uses a single PPO to personalize the entire prompts without dividing them into four patterns, while PPO-4 personalizes the four patterns uniformly with one PPO. The superior performance of RPP compared to PPO-1 and PPO-4 highlights the need for multiple agents to personalize each of the four patterns respectively.}

\subsection{Sensitivity Analysis (\textbf{RQ4})}
We investigate the effect of the number of training examples (users) on RPP/RPP+ with LLaMa2-7B-chat in Figure~\ref{Fig.sub.1}. As the number of training examples increases from $100$ to $500$, the performance of RPP/RPP+ initially improves and subsequently stabilizes. Considering both the performance and resource consumption, we set the number of training examples as 200. Additionally, we explore the sensitivity of RPP/RPP+ to the number of candidate items within the range of $\{5, 10, 15, 20\}$ in Figure~\ref{Fig.sub.2}. Observing the curves corresponding to the task-wise prompting method (\eg enumeration) and our framework RPP/RPP+, the ranking performance of LLMs drops as the number of candidate items increases, which can be attributed to the limitation of LLMs in processing long inputs. However, the capability of RPP/RPP+ to enhance the recommendation performance of LLMs is robust since the gap between the curves of RPP/RPP+ and Enumeration remains stable. 
This observation highlights the effectiveness and stability of personalizing instance-wise prompts rather than solely relying on task-wise prompts.

{Furthermore, to evaluate the impacts of different numbers of sentences under each pattern, we conduct experiments for the four patterns with results presented in Figure \ref{Fig.num_action}. The curves rise initially and then stabilize, demonstrating that when the sentence actions encompass a sufficient range of perspectives for each pattern, the performance of prompt personalization can be sustained at a high level.}

\begin{figure*}[t]

    \centering
    \includegraphics[width=0.75\linewidth]{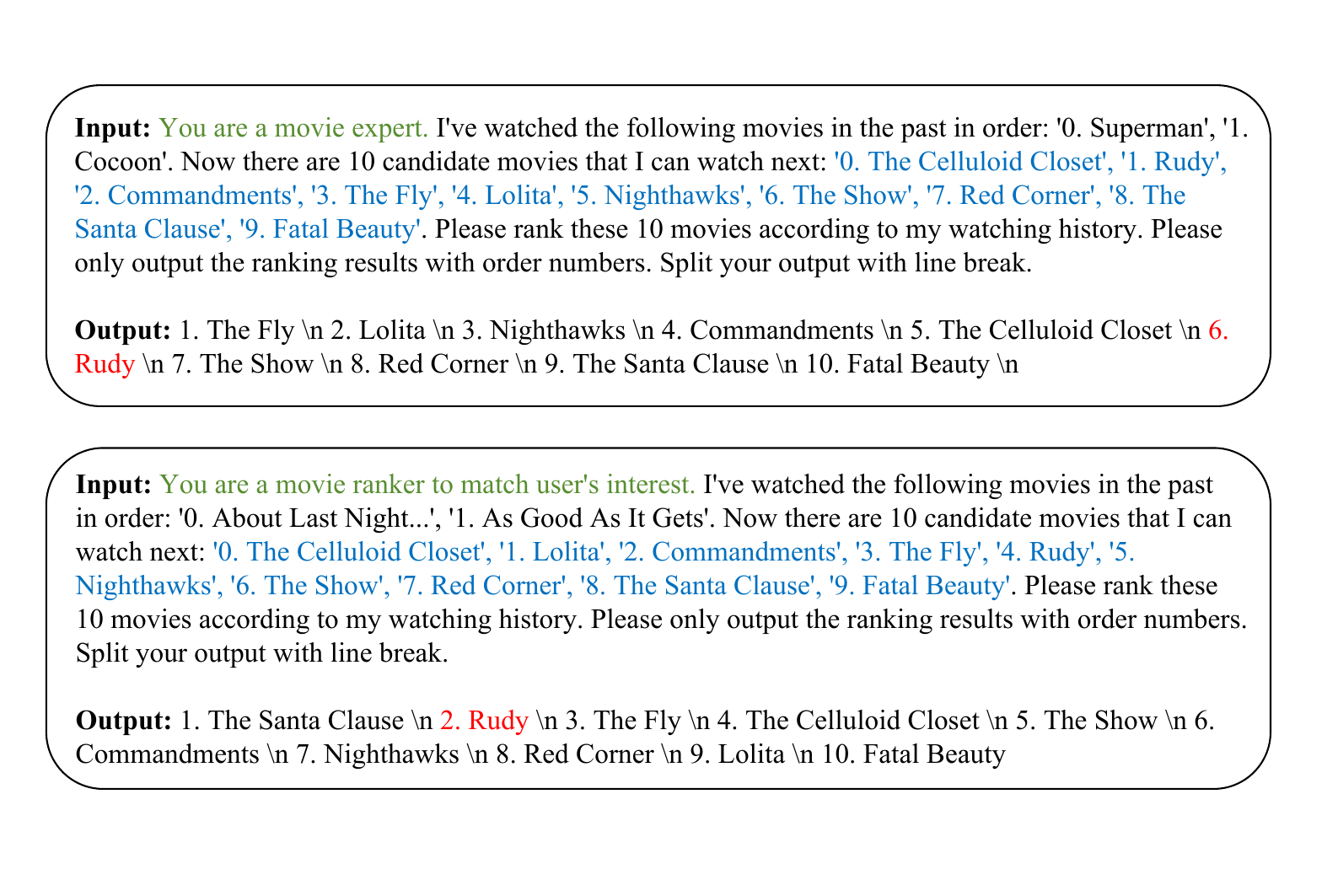}
    \vspace{-6mm}
    \caption{The case study on ``role-playing'' patterns. Variations of prompt pairs are marked {\color{green}green}, the candidate items are marked {\color{blue}blue}, and the ground truth is marked {\color{red}red}.}
    \label{fig:case1}
   
\end{figure*}

\begin{figure*}[t]
    \centering
    \includegraphics[width=0.75\linewidth]{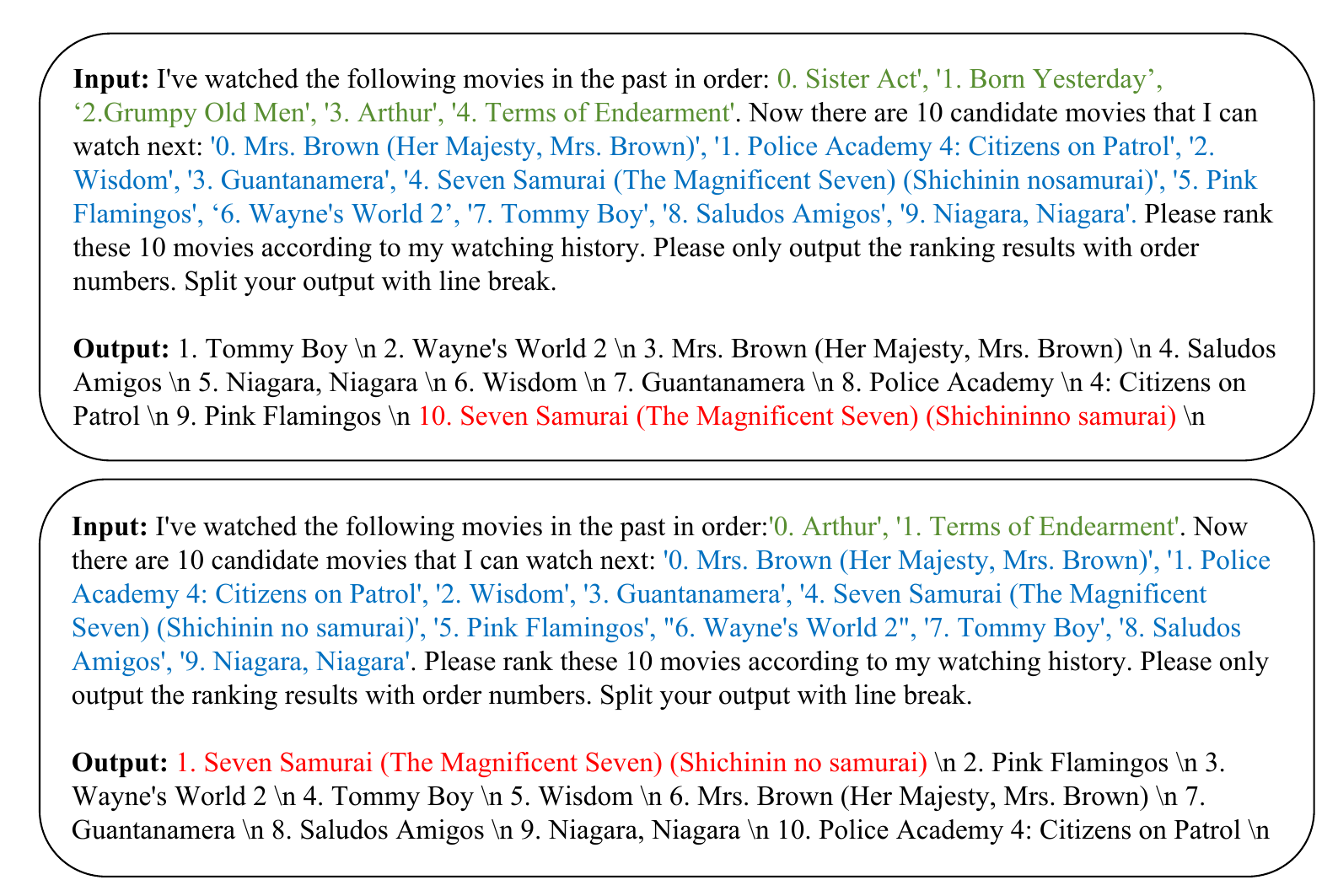}
    % \vspace{-2mm}
    \caption{The case study on ``history records'' patterns. Variations of prompt pairs are marked {\color{green}green}, the candidate items are marked {\color{blue}blue}, and the ground truth is marked {\color{red}red}.}
    \label{fig:case2}
    % \vspace{-2mm}
\end{figure*}

\begin{figure*}[t]
    \centering
    \includegraphics[width=0.75\linewidth]{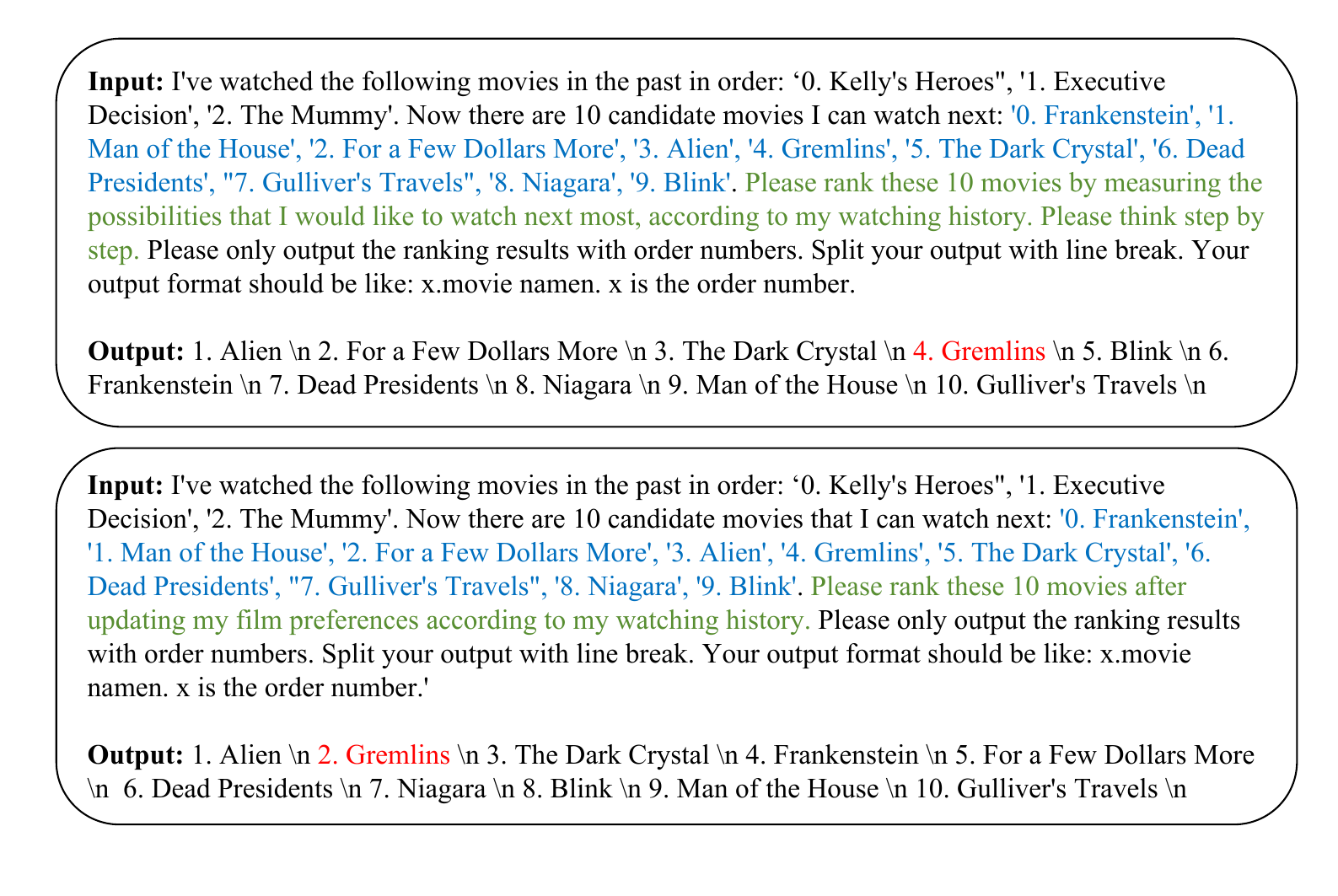}
    \vspace{-2mm}
    \caption{The case study on ``reasoning guidance'' patterns. Variations of prompt pairs are marked {\color{green}green}, the candidate items are marked {\color{blue}blue}, and the ground truth is marked {\color{red}red}.}
    \label{fig:case3}
    % \vspace{-4mm}
\end{figure*}

\begin{figure*}[t]
    \centering
    \includegraphics[width=0.75\linewidth]{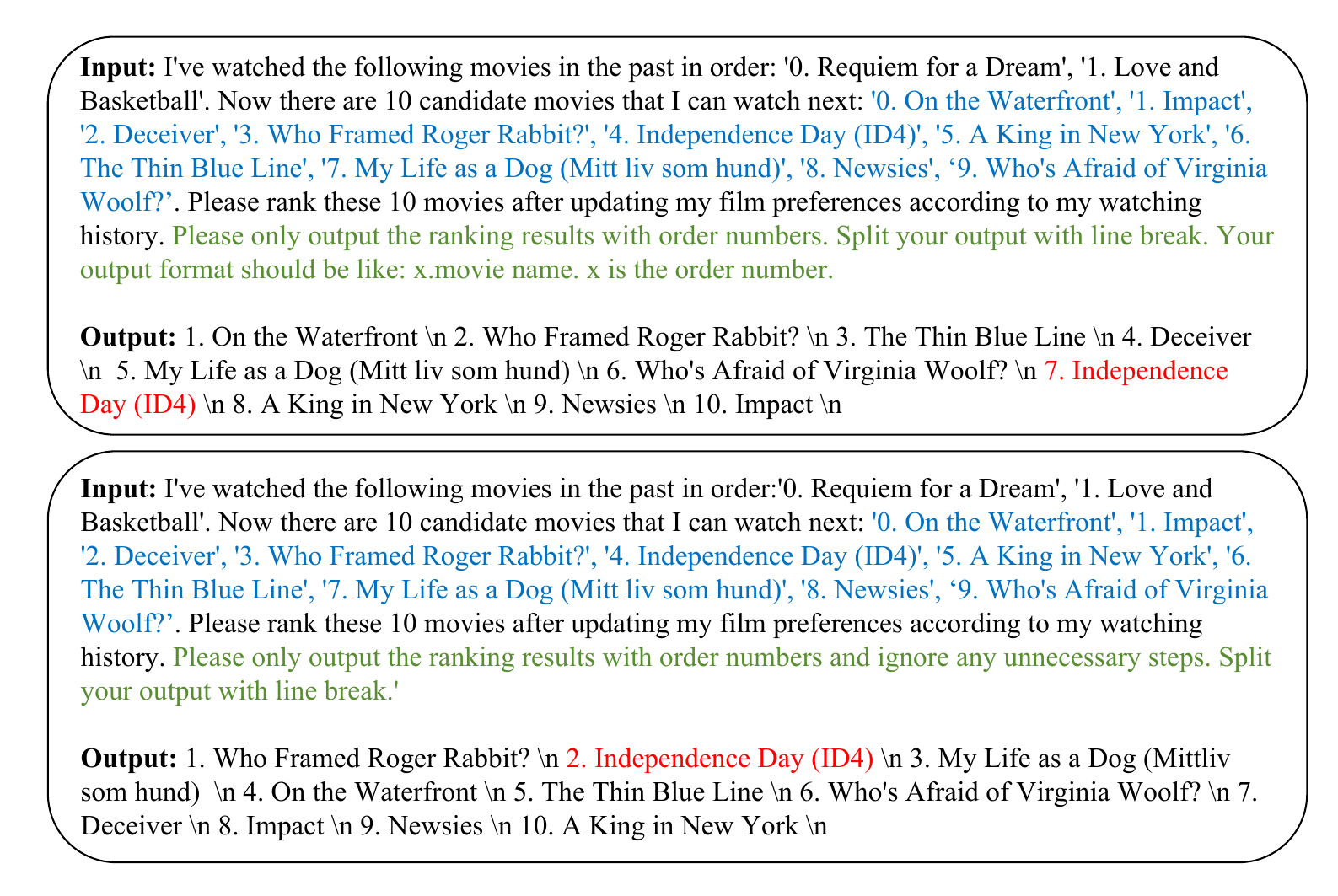}
    % \vspace{-2mm}
    \caption{The case study on ``output format'' patterns. Variations of prompt pairs are marked {\color{green}green}, the candidate items are marked {\color{blue}blue}, and the ground truth is marked {\color{red}red}.}
    \label{fig:case4}
    % \vspace{-4mm}
\end{figure*}

\begin{figure*}[t]
    \centering
    % \vspace{-5mm}
    \includegraphics[width=0.8\textwidth]{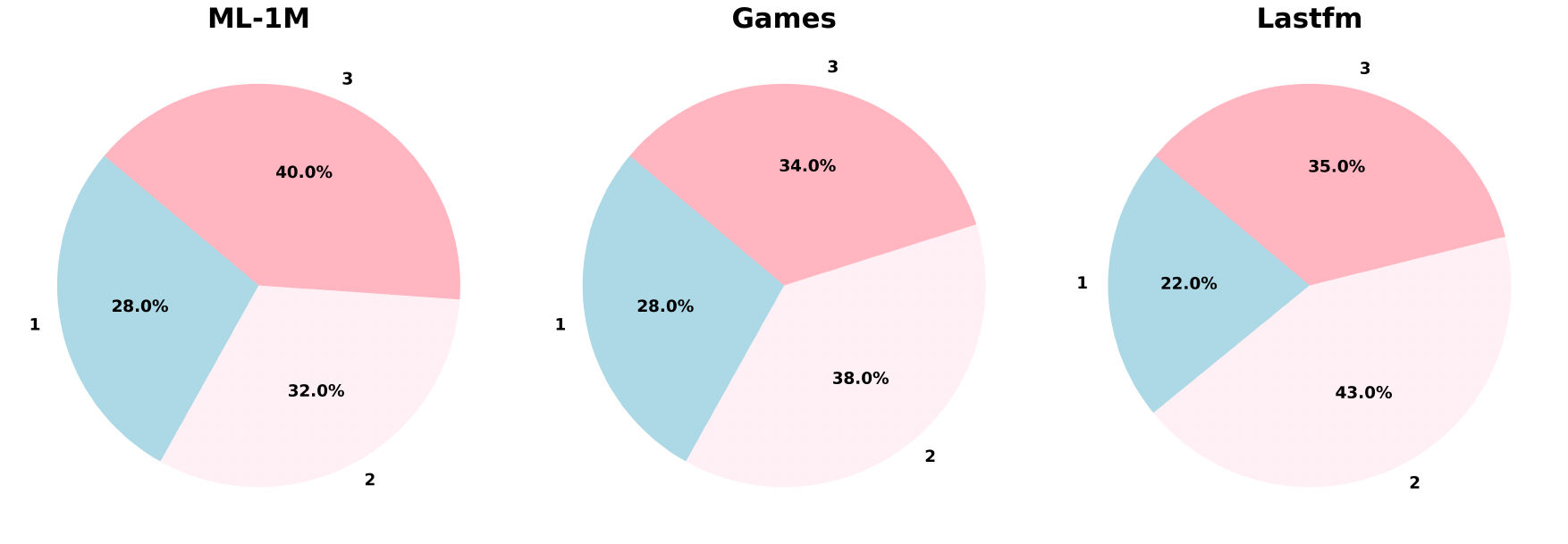}
     \vspace{-5mm}
    \caption{The user distribution on personalized ``role-playing'' pattern, with three sectors representing the three expressions for this pattern.}
    \label{Fig.persona}
    % \vspace{-2mm}
\end{figure*}
\begin{figure*}[t]
    \centering
    % \vspace{-5mm}
    \includegraphics[width=0.8\textwidth]{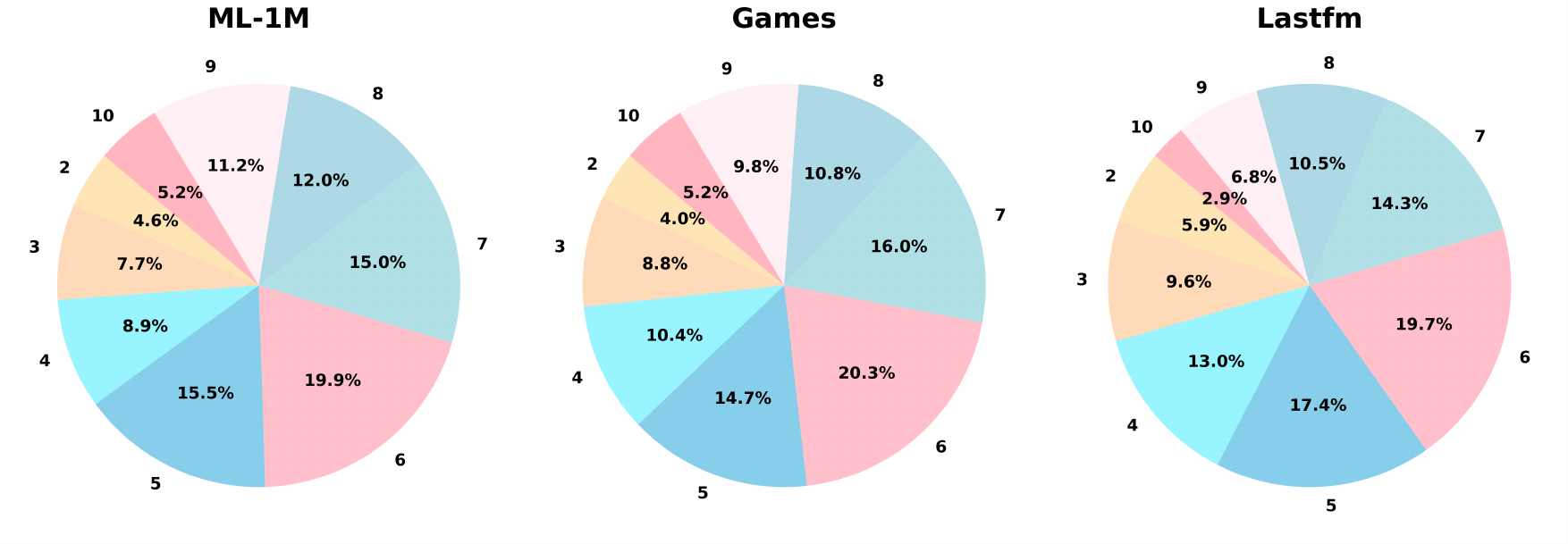}
     \vspace{-5mm}
    \caption{The user distribution on personalized ``history records'' pattern, with each sector representing varied interaction history length.}
    \label{Fig.history}
    % \vspace{-2mm}
\end{figure*}

\begin{figure*}[t]
    \centering
    % \vspace{-5mm}
    \includegraphics[width=0.8\textwidth]{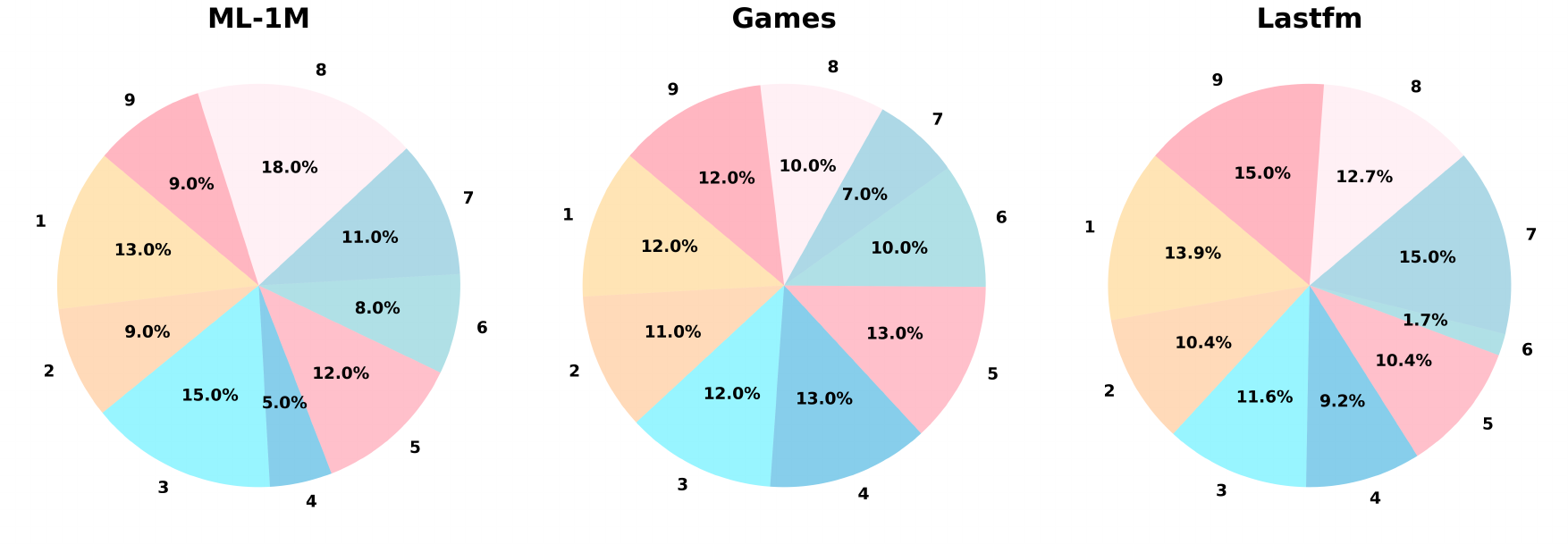}
     \vspace{-5mm}
    \caption{The user distribution on personalized ``reasoning guidance'' pattern, with nine sectors representing the nine expressions for this pattern.}
    \label{Fig.recipe}
    % \vspace{-2mm}
\end{figure*}

\begin{figure*}[t]
    \centering
    % \vspace{-5mm}
    \includegraphics[width=0.8\textwidth]{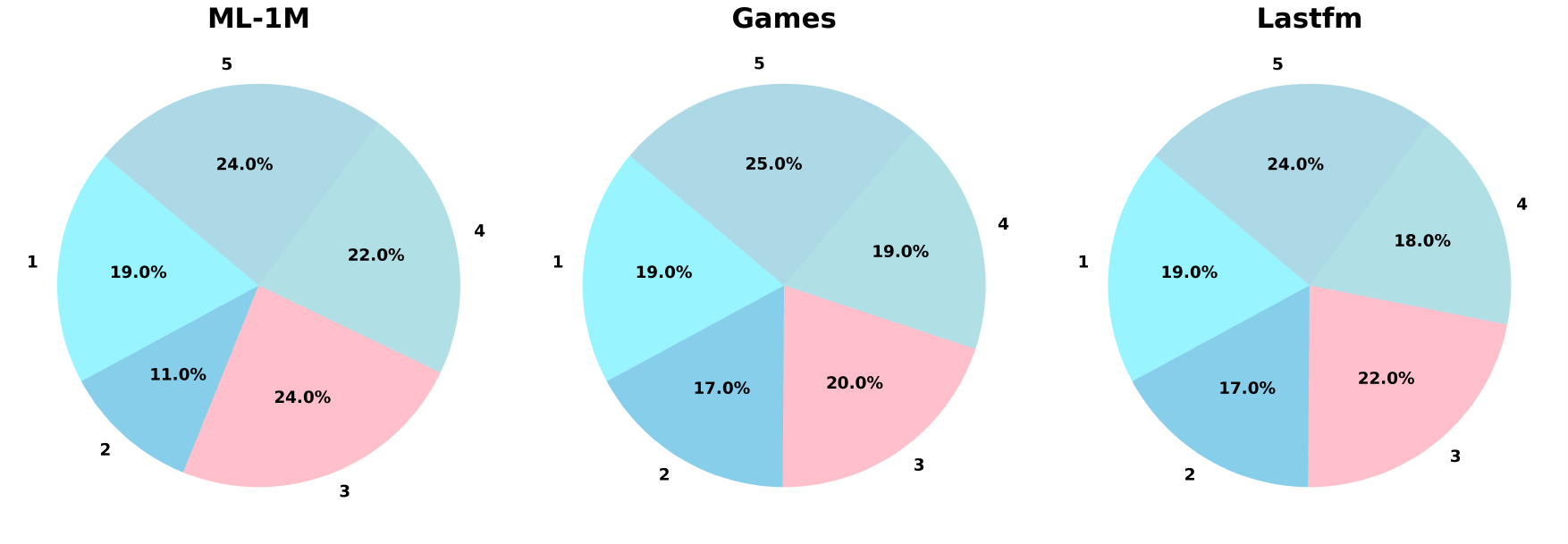}
     \vspace{-5mm}
    \caption{The user distribution on personalized ``output format'' pattern, with five sectors representing the five expressions for this pattern.}
    \label{Fig.template}
    % \vspace{-2mm}
\end{figure*}

\subsection{Case Study}
Through the case study presented in Figure~\ref{fig:case1},~\ref{fig:case2},~\ref{fig:case3}, and~\ref{fig:case4}, we provide an intuitive demonstration of how variations in these prompts impact the ranking results. 
For the user in Figure~\ref{fig:case1}, personalizing the ``role-playing'' pattern from ``You are a movie expert'' to `` You are a movie ranker to match user's interest'' results in a higher ranking of the ground truth item that the user interacts with next. This illustrates that assigning LLMs as movie rankers to analyze users' preferences can lead to more precise recommendations for this user.
By personalizing the 'history record' pattern from 5 interaction records to 2 historical records, the ranking results become more precise for the user in Figure~\ref{fig:case2}. This demonstrates that considering a shorter-term interaction history is more effective in capturing his preferences than a long-term one. For the user in Figure~\ref{fig:case3}, personalizing the ``reasoning guidance'' pattern from ``think step by step'' to `` update the film preferences before ranking'' leads to better recommendations, proving that utilizing LLMs' refinement and update mechanism is more reasonable than deducing without specific instructions for him. Moreover, personalizing the ``output format'' pattern from ``giving the output format example'' to ``answering without any unnecessary steps'' makes it easier for LLMs to understand the instructions and recommend for the user in Figure~\ref{fig:case4}.
We can observe that due to LLMs' sensitivity to prompts and users' inner differences, personalizing prompts for each user enables LLMs to generate recommendation results that better satisfy the users, with the ground truth prioritized. Furthermore, we visualize the distribution of users across different patterns under our experiments' setting to illustrate user diversity vividly, as depicted in Figure~\ref{Fig.persona},~\ref{Fig.history},~\ref{Fig.recipe}, and~\ref{Fig.template}, underscoring the necessity of personalizing each pattern in the prompts for different users. 
This further validates the significance of instance-wise prompting for recommendation tasks and the effectiveness of our framework.

\subsection{Timing and Computational Complexity Analysis}
Compared to directly applying fixed prompts, the streamlined architecture of MARL, with parameters negligible relative to LLMs, incurs minimal additional computational and time costs. When customizing the prompt for each user, compared with heuristical crafting methods like enumeration or token-level optimization approaches like RLPrompt~\cite{deng2022rlprompt} and AutoPrompt~\cite{shin2020autoprompt}, the sentence-level optimization can improve searching efficiency and obtain optimal prompts with minimal iterations, thus reducing timing and computational complexity. However, the multi-round iteration with LLMs is still required since achieving the desired prompt in a single step with MARL is challenging. Formally, let $N$ denote the number of users, $L$ denote the length of a prompt, $D$ denote the possible tokens, and $S$ demote the limited actions. RPP/RPP+ reduces the timing and computational complexity of designing prompts from $O(NLD)$ to $O(NS)$. Our experiments reveal that ChatGPT can complete prompt optimization in approximately three rounds, whereas LLaMa2-7B-chat requires two, seven, and eight iterations on three distinct datasets (\ie{ML-1M, Games, and Lastfm}), and Alpaca takes around 8 iterations. The timing and computational complexity are acceptable for achieving significant performance improvements. It reveals that RPP and RPP+ can achieve a balance between prompt quality and resource efficiency.

%% file: 2_related_work.tex
\section{Related Work}
Here, we provide a review regarding LLMs for recommender systems, prompt sensitivity of LLMs, and discrete prompt engineering.
\subsection{LLMs for Recommender Systems} Recently, large language models (LLMs) have demonstrated remarkable capabilities in instruction following and intent reasoning, drawing researchers' attention to using LLMs as recommender systems \cite{tois_Wu,cirs_gao,FCS_wu,mao2025distinguishedquantizedguidancediffusionbased} which can capture user preference and recommend for them. Conceptualizing LLMs as recommender systems to generate recommendations relies heavily on the quality of the prompts designed. Tailoring prompts of LLMs for the recommendation tasks has been explored by recent research.
InstructRec~\cite{zhang2023recommendation} develops a general instruction format that contains users' preferences, intentions, task requirements, and contextual information. 
Chat-REC~\cite{gao2023chat} recasts user profiles and prior interactions into prompts, enhancing the interactivity and explainability of the recommendation process. LLM4RS~\cite{dai2023uncovering} designs domain-specific prompt format to analyze ChatGPT's recommendation ability on all three ranking policies, including point-wise, pair-wise, and list-wise ranking. Instead of designing a common prompt format for the specific recommendation task, we personalize instance-wise prompt for each user to explore users' different preferences. {Though existing studies \cite{PepRec, Re2LLM, ISR} have explored optimizing prompts for users, they may be limited in their reliance on LLMs to refine or partial optimization from limited perspectives. For example, Peprec \cite{PepRec} refines the initial prompts by incorporating interactions
from similar users in the same cluster iteratively. Re2LLM \cite{Re2LLM} and ISR \cite{ISR} correct LLMs' recommendation errors through self-reflection and reasoning. Different from them, our method decomposes the completed prompt into multiple key patterns and optimizes each pattern comprehensively with reinforcement learning.}

\subsection{Prompt Sensitivity of LLMs}
Recent studies~\cite{lu2021fantastically} have examined the sensitivity of LLMs to prompts that even minor variations in prompts can lead to significantly different outputs, indicating that the one-size-fits-all prompt may not generalize well \cite{liu2023promotinggeneralizationexactsolvers,InfoIGL_FCS,tkdd_ood} for different inputs. For example, KATE~\cite{liu2021makes} confirmed that the number and order of in-context examples can influence the output of LLMs. Besides, variations in prompt formats within task descriptions may result in LLMs interpreting input instances differently, consequently influencing the outcomes~\cite{zhao2021calibrate}.
These studies underscore the importance of optimizing prompts for specific inputs to enhance the performance of LLMs. Fixed prompt templates that assume uniformity across all samples in a task, regardless of their varying difficulty, may impair LLMs' abilities due to a lack of instance-specific knowledge~\cite{jiang2022instance}. To address the difference in inputs, recent research has proposed instance-level prompt optimization~\cite{jin2023instance}, which involves rewriting the prompt for each input to better leverage the LLM's capabilities for specific instances. IPL~\cite{jin2023instance} assumed that each learnable prompt token contributes differently to various instances, which can be learned by calculating relevance scores. To meet the specific demands for personalization in recommendation tasks, we enhance LLMs' personalized recommendations by leveraging instance-wise prompts and introducing a method to optimize these prompts using MARL.

\subsection{Discrete Prompt Engineering}
Several approaches have explored prompt engineering in continuous and discrete space to obtain better answers from LLMs. Since the prompts for recent LLMs (ChatGPT, LLaMa2) are discrete and hard prompts, our work is strongly related to the discrete prompts engineering.
Prior works have attempted to construct discrete prompts manually~\cite{petroni2019language} or by heuristic~\cite{jiang2020can}, which are limited by human effort or resource consumption. Recent research has explored constructing prompts automatically. For instance, prompt retrieval~\cite{rubin2021learning} leverages annotated data and a language model, training a compact retriever to retrieve prompts based on input-output pairs. AutoPrompt~\cite{shin2020autoprompt} utilizes gradient-guided search to identify the optimal tokens within the prompt, although these prompts are typically not human-interpretable. RLPrompt~\cite{deng2022rlprompt} introduces a framework based on RL to generate prompts iteratively word-by-word from the vast vocabulary.   
An important distinction between our method and the existing methods is that we decompose the completed prompt into multiple key patterns, optimizing each pattern and concatenating them, which can enhance the efficiency of optimization and ensure the quality of prompts. To the best of our knowledge, our work is the first to explore instance-wise prompting in the domain of RSs.

%% file: 6_conclusion.tex
\section{Potential Application and Limitations}
In real-world applications, after the initial coarse ranking process, RPP/RPP+ can be utilized in the fine-ranking stage to deliver highly relevant and personalized recommendation results to users based on their preferences and interests. By employing personalized prompts, LLMs can conduct more refined ranking, leading to a more accurate understanding of users' underlying preferences. Additionally, the personalized interaction history length can capture the dynamic changes in users' interests within sequence recommendation scenarios. Furthermore, the personalized reasoning process, as realized by the pattern of ``recipe'', can improve the interpretability of recommendations.

While the proposed RPP/RPP+ framework presents an efficient solution for personalizing instance-wise prompts for recommendation, it is accompanied by certain limitations.
A notable limitation is its partial reliance on manual intervention when designing the action set in the initial setup phase. This suggests a potential area for future research to develop more autonomous and adaptive methods.
Another limitation is the efficiency of RPP/RPP+ in narrowing the search space through iterative updates, which necessitates a certain level of computational resources. In the future, we will explore a more efficient method for prompt personalization, preferably without relying on LLMs in the iterations.
\section{Conclusion}
In this paper, we propose instance-wise prompting to personalize prompts for individual users in recommendation tasks. To solve this task, we introduce a new framework, Reinforced
Prompt Personalization (RPP/RPP+), personalizing essential patterns respectively with multi-agent and then concatenating them into a personalized prompt. We optimize the prompts at the sentence level and carefully design the expressions for each pattern, enhancing search efficiency and ensuring prompt quality.
Experimental results validate the superiority of RPP/RPP+ over several traditional recommender models, few-shot methods, and prompt-based methods in ranking tasks, demonstrating the potential of personalizing instance-wise prompts. 

\section{Acknowledgements}
This research is supported by the National Natural Science Foundation of China (No.92270114, No.62302321, No.U24B20180) and the advanced computing resources provided by the Supercomputing Center of the USTC.